\documentstyle{article}
\begin{document}
\begin{center}
{\bf NGC 4151 - A Unified Active Galactic Nucleus\\}         
I. Cassidy \& D.J. Raine \\                                      
Department of Physics and Astronomy, University of Leicester \\
Leicester, LE1 7RH, UK\\
21 Nov 1995; revised 17 July 1996\\
Key-word numbers: 11.01.2, 11.09.1, 11.17.3
\end{center}

\begin{abstract}        
We present a unified picture of active galactic nuclei which we construct from a
detailed model of line emission in the active source in NGC 4151.
This source provides us with
an opportunity to explore the variation of structure 
with luminosity in an object which,
in the model, derives its unusual properties as a consequence of the
angle of the accretion disc to the line of sight. 
The key features that emerge from the model are (i) a
non-spherical broad line region (BLR); (ii) short-lived BLR clouds; (iii) a 
luminosity-dependent
structure for the BLR; (iv) a luminosity-dependent flared accretion disc 
extending beyond the BLR and (v) a
separate intermediate line region between the BLR and NLR.  
The special orientation turns out to enable us to
fix many of the parameters of the model for this source. It is then natural
to ask how
this model would appear, in its various luminosity states, 
at other orientations. To make contact with observations we need to include
dust obscuration over a range of angles near to the plane of the disc. 
We then obtain the spread of observed types of radio-quiet
active  nuclei
and we propose an extension of the model to a 
unification of radio-loud active
galaxies. Thus, in this scheme  NGC 4151
can be regarded as a typical active nucleus, special only in its orientation.
We shall find that this alleviates a 
number of problems with a unified picture based on a dusty
molecular torus with a fixed structure (to which NGC 4151 appears as an
anomaly). In the proper sense of
the expression, NGC 4151 may be the exception that proves the rule.
\end{abstract}                                 
\section{Introduction}

In order to explain polarisation data for the
radio galaxy 3C 234  Antonucci (1984) proposed a model in which the 
continuum source and broad-line clouds
are located inside a thick torus, which blocks these from direct
view, while electrons above and below the source
scatter continuum and broad-line photons into the line of sight.
This picture was extended following the discovery that the
type 2 Seyfert galaxy NGC 1068 harbours a nucleus, of the same
nature as a typical Seyfert type 1 nucleus, which is hidden from
direct view but which is revealed indirectly
in scattered light.
A model thus evolved of a partially obscuring dusty
torus as a universal feature of Seyfert galaxies, in which observed 
differences are supposed to arise only from different lines of sight to
intrinsically similar objects. 
The thickness of the torus and
its opening angle are assumed to be independent of luminosity. Indeed, in the
extreme hypothesis that Antonucci (1993) calls the straw person's model (SPM), 
the only difference amongst AGN is the division into radio-quiet and 
radio-loud
types. The opening angle of the torus is then set by the observation of
polarisation perpendicular to the radio axis in NGC 1068. Since a scattered
photon retains a polarisation orthogonal to the propagation vector prior to
scattering the bundle of ray directions before scattering cannot
diverge too much. This gives a half-opening angle of at most 30$^{\circ}$,
consistent with the ratio of Seyfert type 1 to type 2 galaxies if the latter
are all obscured views of the former. 

But in schemes such as the SPM in which the structure proposed for
NGC 1068 is typical, the active nucleus of 
NGC 4151 presents itself as a unique exception,
essentially because the line of sight over the putative obscuring
torus is incompatible with the view of the observed ionisation cone
in the extended narrow line region (Penston et al. 1990, Evans et al. 1993)
and possibly also the
extended linear radio structure 
along the axis of the torus  (Antonnuci 1993).

The variation on the theme of unification that 
we present in this paper
is, in contrast, based on a detailed analysis  of NGC 4151. Our picture of 
this galaxy,  which results from fitting emission
line properties, is described in Cassidy \& Raine (1996) and
summarised in section 2. 
We  make the assumptions (i) that the
accretion disc extends  out to the BLR and (ii) that there is a supersonic 
nuclear
wind generated close to the black hole. 
The BLR clouds 
arise from material injected from the accretion disc into the outflowing
wind (Cassidy \& Raine 1993). 
In the supersonic flow the BLR clouds are short-lived. 
Therefore the BLR is confined to a region close to the disc surface. 
This means that 
the BLR can respond rapidly to changes in the continuum, in contrast to a 
spherical BLR (section 5.1). Thus if changes in the continuum result in 
a loss of  BLR clouds, as we showed previously,  a change of Seyfert type 
can occur on short timescales. 

The dispersing cloud material provides a scattering medium by which the BLR and
UV/optical continuum
may be observed in systems where it is obscured from direct view. From other
lines of sight this same material is the X-ray warm absorber. To account for
the obscured systems we introduce a dusty torus in the region
between the BLR and NLR, which may also be
responsible for at least part of the IR emission (Efstathiou \& Rowan-Robinson 
1995).
The non-spherical BLR geometry (and the location of
the scattering medium) turns out to imply that the torus does not have to have a 
{\em small} opening angle. The existence of  Seyfert 2 galaxies truly lacking 
a broad-line region,
predicted by the model, takes care of the ratio of Seyfert types, which would
otherwise be increased in favour of Seyfert 1s by the larger opening angle. 
This also provides a possible solution to the
over-production of the X-ray background, because the number of obscured
high luminosity Seyfert 1s is reduced.  

The model itself determines
differences in systems at lower and higher luminosities than the range covered
in NGC 4151 and this is shown to account for the whole of
the radio-quiet class of AGN. In particular, the obscuring torus is here part 
of a flared accretion disc
and is located in a
region where it may be influenced by the nuclear properties such that the 
opening angle increases with luminosity. 

The following list of contents of the paper will serve to explain its
structure. Section 2 reviews the basic features
of our model for NGC 4151. In section 2.1 we argue that the material from
the dispersing broad-line clouds produces a medium that can provide a
significant electron scattering optical depth to the continuum and broad-line
emission. In section 2.2 we show that, for other lines of sight, 
the dispersing cloud material also provides a warm absorber in the X-ray band.
For NGC 4151 we show that the line of sight implies a combination of partial
covering of the X-ray source by cold clouds and by this warm absorber, and
show this prediction is consistent with Ginga X-ray data. We argue that the 
UV absorption
may be explained without invoking any additional components in section 2.3.
Changes in the continuum illumination of the disc lead to suppression of the
broad-line
cloud production and in section 3 we argue that this is consonant with the 
observed
changes in broad-line properties in intermediate Seyferts. We predict the
existence of true Seyfert 2 galaxies, lacking a broad-line region. 
In section 4
we argue that the `unified Seyfert' must contain an obscuring dusty torus, 
and that
the presence of such a torus is consistent with the properties of NGC 4151
provided that the opening angle is sufficiently large. Having established the
structure of NGC 4151 we investigate what it would look like at different
angles. The importance of a flattened cloud geometry for the BLR is discussed 
in section 5.1. The X-ray absorption is considered as a function of the 
angle of the disc to the line of sight in section 5.2 
The scattering into the line of sight of the obscured inner region
at low inclinations is considered in section 5.3. Arguments for the presence of
an ILR in all higher luminosity AGN are considered in section 5.4. 

Section 6.1 deals with the flaring of the
disc as a function of luminosity. We explain here why we consider the 
opening angle
of the obscuring torus to be luminosity dependent. As a result, at low
inclinations in high luminosity systems we predict broad absorption line
quasars and explain why we believe this type of system will be radio-quiet.
In the next sections we extend the model to lower luminosities (section 6.2) 
and higher luminosities (section 6.3).
Finally, in section 7, we 
discuss briefly how the model
extends rather naturally to accommodate radio-loud objects in a 
grand-unified picture.

\section{The structure of NGC 4151}
 
Our model for the nucleus of NGC 4151 (Cassidy \& Raine 1996)
is sketched in figure 1 and the main parameters listed in table 1. 
There are three emission line regions: the broad (BLR),
intermediate (ILR) and narrow (NLR) line regions. The BLR consists of
clouds injected from the surface of an externally heated accretion disc 
(as a result of
Kelvin-Helmholtz instabilities)
into a supersonic radial outflow, where they are accelerated and destroyed by
ram pressure.  The parameters
of the system (the luminosity of the central ionising source, $L_{r}$, 
the black hole mass, $M_{bh}$,                   
and the momentum flux in the wind, $L_{m}$) determine the inner and outer radii
of the BLR, $r_{in}$ and $r_{out}$, the
flaring of the disc in the BLR region and
the cloud density and column density. The injection velocity of the clouds 
at the disc surface, ${\bf v}_{in}$, has a normal component $v_{n}$ 
of the order of the sound speed at the 
Compton temperature of the radiation field falling on the disc; the
azimuthal component is taken to be the Kepler velocity at the injection radius.
Part of the observed line emission comes from radiation reflected from the disc
surface. The disc is opaque in the BLR and therefore obscures clouds on the far
side. Assuming a value for the 
albedo of the disc, $A_{d}$, enables the line profiles from the
cloud emission to be computed as a function of viewing angle for a specified
ionising continuum. The three components of the cloud velocities are computed
assuming the cloud dynamics is governed by gravity and ram pressure
acceleration and that a cloud loses most (90 \%) 
of its mass through the flanks over 
a time, $t_{c}$, of the order of the sound crossing time, $t_{s}$ in the
cloud at formation.
For definiteness, we fix this as $t_{c}=2t_{s}$. The sound crossing time is
short compared with the crossing time of the BLR at the cloud speed, so the BLR
geometry is disc-like. The reflected component of the line emission is Doppler
shifted by the rotation of the disc, which is taken to be Keplerian.

The luminosity of the source is estimated from observation and the
parameters $M_{bh}$, $L_{m}$, $v_{n}$ and $A_{d}$  and the viewing angle
are chosen to fit the profile of the               
CIV $\lambda 1459$ line when the nucleus is in its high luminosity
state (Fahey et al. 1991).
These parameters appear to be fairly well
constrained {\em by fitting the CIV line profile alone} given the cloud survival
time.  
The density in the innermost clouds, $n \sim 10^{12} $ cm$^{-3}$
gives an ionisation parameter for $r_{in}=10^{16}$ cm consistent with
the line ratios, this
small BLR size being confirmed by the line variability over 8-16 days in CIV 
and Ly$\alpha$ (Clavel et al. 1991) and over 9$\pm2$ days in H$\beta$ (Maoz et
al. 1991).
The BLR is stratified ($n\propto 1/r^{2}$) 
and the run of density
is consistent with a relatively weak CIII] $\lambda$ 1909 line. For a Mathews
and Ferland (1987) spectrum we find general agreement for the usual diagnostic
line ratios (see Table 2 in Cassidy \& Raine, 1993). 
We find $M_{bh}=3\times 10^{7} M_{\odot}$,
$r_{in}=1.4 \times 10^{16}$ cm and $r_{out}=10^{17}$ cm, $v_{n}=
2\times 10^{8}$ cm s$^{-1}$; also, $L_{m}/L_{E} \leq 5$, 
where $L_{E}$ is the
Eddington limit luminosity, and $A_{d}=0.3.$
Under the combined effects of surface heating by the
external radiation and the action of
the ram pressure of the nuclear wind, the disc 
flares to an angle of about $30^{\circ}$ as required to fit the kurtosis
(width at half maximum/width at zero intensity) 
of the CIV line. The angle of the line of 
sight to the disc is then $32^{\circ}$. For fixed values of the other  
parameters this angle is constrained to 
within at most a few degrees by fitting
the profile of CIV by eye. (This is because line profile changes are quite marked 
as the line of sight approaches the angle of the flared disc.)
The presence of line emitting  
material close to the line of sight to the continuum source 
is consistent with  the observed response of the lines to changes in the 
ionising continuum with a delay of less than 3 days (Clavel et al. 1991)
and this is confirmed by the 
non-zero response at zero continuum lag in the 1D transfer function (i.e. the
line response to a $\delta$-function ionising continuum (Blandford \& McKee
1982)). Details of the transfer functions obtained by Horne \& Ulrich (1996) 
can also be accounted for (see section 5.1). Further details are given in 
Cassidy and Raine (1996). 

As it declines in luminosity NGC 4151 changes its classification from
a Seyfert 1 to close to a Seyfert 2. In this transition 
the shapes of the emission and absorption lines change, so this is not 
just an effect of the loss of ionising photons. The lifetime of the BLR clouds
is $10^{4}r_{16}^{2}$ s, so rapid changes in structure are possible. From the
low state profiles we deduce the existence of the ILR lying in the range 
$3\times 10^{17}$cm $<r<10^{19}$ cm. (See also Brotherton et al. 1994.) 

We have suggested (Cassidy \& Raine, 1996) 
that the ILR arises from a thermally driven disc wind (compare
Smith \& Raine, 1985), hence
that it too has a flattened geometry. We have shown also that, in the low
state, the inner BLR
clouds disappear, enabling us to explain
the changes in the
CIV profile width and equivalent width.
Other optical and UV line ratios and profile characteristics
in the various luminosity states can also be accounted for.

\subsection{Debris from the BLR clouds}
The warm material from the dispersing BLR clouds forms a scattering medium. 
The optical depth can be estimated as follows. 
Let the clouds rise from the disc over an area $A$ and let 
the mass input into the clouds be $\dot{M}$.
Suppose the heating timescale for the 
cloud debris is $t_{h}$. The column of warm dispersing material is                 
\[
N_{h} \sim \dot{M}t_{h}/m_{H}A.
\]
We have $t_{h}=kT/\Gamma$, where $\Gamma=(\sigma_{T}/m_{e}c^{2})(L/4\pi
r^{2})4kT_{c}$ is the Compton heating rate for a Compton temperature $T_{c}$. 
This
gives $t_{h}\sim 10^{8}r^{2}_{17}(L/L_{E})m_{8}$ s at $10^{17}r_{17}$ cm.
 We estimate the mass flow rate into the BLR of NGC 4151 by comparing the 
model line flux for CIV $\lambda 1549$ 
with the observed flux in the high luminosity state. For a distance to 
NGC 4151 of 20 Mpc we obtain 
a mass flux of 
2.4$M_{\odot}$ yr$^{-1}$.
Then, with ${\rm d}\dot{M}/{\rm d}r \propto 1/r$, we find that  
the dispersing column has
an electron scattering optical depth of  $0.1 (L/L_{E})m_{8}.$ 

The cloud velocity on break-up is much less than the wind velocity, $v_{w}$. 
The acceleration timescale for the dispersing material of thickness $h$,
density $n_{c}$ and temperature $T_{c}$
to reach the ambient wind velocity is $(n_{c}/n_{w})(h/v_{w}).$ Taking for
$n_{c}h$ an upper limit for the  
column density of $10^{23}$cm$^{-3}$, as in the clouds, and putting $n_{w}
=4\times 10^{6}(L_{m}/L_{E})r_{16}^{-2}$ cm$^{-3}$ from Cassidy and Raine
(1996) gives an acceleration timescale $ \leq
2 \times 10^{8} (10^{4}{\rm K}/T_{c})r_{17}^{2}$ s, of order of the heating 
timescale. 
Thus, once the material has started heating it is rapidly dispersed. In other
words, most of the debris 
is in material heating to $10^{5}$K
and moving with the clouds. It will therefore contribute a scattered
broad-line component that is polarised
transverse to the disc axis and which is not significantly broader 
than the unscattered flux. 
Note that the scattering into the line of sight that
has been observed from much more extended regions (section 5.3), and probably
dominates in NGC 1068, must have a different explanation (Capetti et al. 1995,
Antonucci et al. 1994). 

\subsection{X-ray properties}
The variability of the X-ray source in NGC 4151 means it must lie 
within 10$^{15}$ cm;
the nuclear wind
is optically thin to electron scattering to this radius.

The dispersing material from the
inner clouds provide warm absorbing material in the line of sight to the
X-ray continuum source and the outer BLR clouds partially cover the source. 
The model suggests therefore that in this object we
have a {\em combination} of warm absorber and partial coverer for the X-rays
when the nucleus is in its high luminosity state. Reynolds and Fabian (1995), 
in discussing MCG 6-30-15, have also pointed to the possibility that the
dispersing BLR material may provide the X-ray warm absorber.
Yaqoob and Warwick (1991) excluded a warm absorber in their analysis, but
Warwick et al. (1995) find that a modest level of
ionisation provides the required decrease in opacity in the 1-4 keV band whilst
maintaining an iron K-edge energy consistent with the average of $\sim 7.3$ keV
derived in the Ginga observation (Yaqoob et al. 1993). Weaver et al. (1994a \& b)
model the ASCA data with either a dual cold absorber plus soft
excess components, or a warm absorber plus a scattered continuum. 

It follows
that limited data can not be used to distinguish between models. Nevertheless
it is of interest to note that 
it is possible to fit the very different Ginga spectra (using XSPEC) 
of Nov 1990 and May
1990 (Yaqoob et al. 1993) within the context of our model. We use a
power law continuum  with a photon index $\Gamma$ plus low energy absorption
due to a uniform column ($N_{H} = 10^{22}$ cm$^{-2}$) of cold gas.
This column could represent the ILR or,
possibly, the outer disc material in the line of sight. The abundances used are
solar, except iron which is fixed at twice solar.
An iron emission line and edge are also included. The iron K-edge and emission
line energies are fixed at 7.1 and 6.4 keV respectively. (Allowing them to vary
does not improve the fit.) The iron line is narrow in both fits 
($\sigma = .1$ and
equivalent width approximately 22eV), but is not very well constrained by the
data. The column density varied from $2 \times 10^{23}$ cm$^{-2}$ in May 1990
to $7.4 \times 10^{22}$ cm$^{-2}$ in Nov. 1990, while the covering factor
varied from .16 to .57. In the former case we found $\Gamma=1.76$ and in the
latter $\Gamma = 1.61.$ The respective reduced $\chi^{2}$ values are .71 and
1.12 for 35 df (Fig. 2). In the earlier observation the warm absorber 
dominates and in the latter the partial coverer. Yaqoob et al. (1993) fit their
data with a partial coverer only, but with an iron abundance that is
required to vary between data sets by a factor of 5. With the inclusion 
of the warm
absorber component we find the iron abundance can be held constant. 

Yaqoob et al. also claim that the dip in the covering factor and the 
increase in
column density {\em rule out a cloud model} for the absorption and suggest 
a `bulk'
structure the geometry of which is at least most of the time stationary
relative to the X-ray source. In their view the edge of an accretion disc or a
torus is an obvious candidate. An explanation in the context of our model could
be that an increased radiation pressure or ram pressure in the wind is
flattening the cloud trajectories giving rise to an increase in column and a
decrease in covering factor of the X-ray source. Our conclusion is therefore
that our model is {\em not} ruled out by current X-ray observations.

The small angle of the line of sight to the disc is consistent with a very weak 
reflected hard X-ray component (Maisack \& Yaqoob 1991) in contrast to
Pedlar et al (1992) who suggested an angle of about 40$^{\circ}$ to the radio
axis. 

\subsection{UV Features}                            
We expect  UV absorption lines to arise in the outer BLR and ILR clouds from
the covering of both the UV continuum source and the inner line emitting     
clouds. The
absorption lines are narrow because the flaring of the disc means that the
outer clouds are moving across the line of sight when they cross the inner
clouds (compare BAL QSOs
below). The absorption will be complex and comparison with observation will
have to await detailed modelling. Bromage et al. (1985) argued that a column of
40 - 180 $\times 10^{21}$ cm$^{-2}$ of gas at
$10^{4}$ - $10^{5}$K could be responsible for both the UV and soft X-ray    
absorption which would be consistent with our proposal. On the other hand
while Ulrich (1988),  
locates the absorbers at $\sim 10^{17}$cm with a density in the range 
$10^{8.5}$ - $10^{10}$ cm$^{-3}$, consistent
with our outer BLR clouds, she finds an EW of the CIV absorption of $4-10 \AA$.
This translates to an equivalent H-column of $10^{19}$ cm$^{-2}$ if all carbon
is ionised to C$^{+++}$ and the lines are optically thin, 
and implies that the UV absorption is occurring in a
separate component. The final assumptions may however be incorrect: a
photoionisation calculation using the  CLOUDY photoionisation code shows that 
there is substantial C$^{++}$ in our optically thick outer BLR clouds. Thus it
remains to be seen whether the outer BLR clouds are responsible for the UV           
absorption.             
 
The disc, broad-line clouds and the cloud debris will absorb 
ionising radiation from the 
central source outside of a cone centred on the disc axis. To obtain the angle
of this cone requires detailed modelling since we need to estimate timescales
to better than a factor 2. The following estimate shows only that the
obscuration by the debris in the UV could be consistent with the observed 
emission cone. The cloud    
material rises above the disc through an angle $ \tan^{-1} v_{i}(t_{c}+t_{a})/R_
{BLR} $, where $t_{a}$ is the acceleration timescale 
over which the trajectories of the debris in the wind become effectively
radial. Estimating the timescales as above gives $t_{c}=N_{col}/(n_{c}c_{s})
\sim 10^{7}$ s at $10^{17}$ cm, for a column of $10^{23}$ cm$^{-2}$ 
and a gas density in
the clouds of $3\times 10^{10}r_{17}^{-2}$ cm$^{-3}$, and $t_{a} \sim 2.5 \times
10^{8} r_{17}^{2}$ s. Thus, $t_{a} \gg t_{c}$ and we can neglect the motion of
the undisrupted cloud. The disc rises to an angle of $30^{\circ}$, a height of 
6$\times$ 10$^{16}$ cm  at $10^{17}$ cm.
So the debris, at its input velocity of $2\times 10^{8}$ cm s$^{-1}$, rises  to
about $\tan^{-1} 1.1 $ or about 50$^{\circ}$.
Thus, the unobscured ionising
cone has a full angle of about 80$^{\circ}$.
Within this cone ionising flux escapes to the NLR.
Evans et al. (1993) give $75\pm 10^{\circ}$ for the full cone angle.
The radio jet is presumably associated 
with collimation of the nuclear outflow along the axis (Smith \& Raine 1985).

\section{Intermediate Seyferts and True Seyfert 2s}
For the purpose of matching profile changes in a single galaxy
the disappearance of the inner BLR clouds can be attributed either to
obscuration or to the suppression of cloud formation. 
These two alternatives will give very different predictions for different
lines of sight if the obscuration is angle dependent. 
Changes in ionisation parameter have also been proposed (Kwan, 1984). 
However, Kwan considered only changes in line ratios; the line {\em profile} 
changes observed in NGC 4151 cannot be explained in our picture by variations 
in ionisation parameter alone. 
We shall now argue 
that observational evidence supports the case that 
the formation of the inner clouds  
is indeed suppressed in variable objects in their low luminosity states.
Some of the distinctive predictions of our unified
picture follow from this conclusion.                                  

Variability observations of some AGN seem to
support the hypothesis that changes in classification 
can be explained by dust clouds crossing
the line of sight. For example,
Fairall 9 ($M_{V}$ = -24), a quasar with a Seyfert 1 
spectrum at the time of discovery (Ward et al. 1978) was by 1984 
a Seyfert galaxy ($M_{V}$ = -21) with a spectrum approaching that of a
Seyfert type 2 (Kollatschny \& Fricke 1985). Most of the change appeared in the
years between 1981 and 1984.                    

However, a number of Seyfert galaxies are 
known to change their classification along the sequence between type 1 and 
type 2 on timescales of weeks to months (Antonucci \& Cohen 1983; Ulrich 
et al. 1984; Wamsteker et al. 1985; Andrillat \& Souffrin 1968), which
makes it unlikely that a dust cloud in a torus moving across the line of sight 
is responsible. 
The broad  H$\beta$ line in NGC 5548 disappeared in less than 
three months during 1992, when the nucleus changed from the usual Seyfert 
type 1.5 
to a type 1.9 (Jijima et al. 1992, Penfold 1992, 1992a). 

Loska et al. (1993) modelled a dust cloud surrounding
the nucleus of NGC 5548, and found that a drop in the luminosity by a factor
of 5 can result in the appearance of dust grains at distances as small as
4 lt-weeks but the time-scale of formation of these new dust grains is far
too long to explain the changes. In addition, it is
difficult to explain why the disappearance of the lines is not accompanied
by a corresponding fall in the continuum luminosity if the line weakening
is due only to the increased dust extinction. They conclude that the observed 
changes are independent of the presence of dust.
 
Similar observations to the ones in NGC 5548 were made by Alloin et al. 
(1985) when monitoring
the nuclear H$\beta$ and H$\alpha$ emission profiles of the Seyfert 
galaxy NGC 1566. The spectrum changed from one closely resembling
a Seyfert type 1.9 to  one resembling a Seyfert type 1.2,
with most of the changes taking place within 4 months.
%
The nonstellar continuum 
component increased by a factor of 5 at 3700 $\AA$ over this period. The Balmer
decrement in both broad and narrow components is low in NGC 1566, implying 
little or no dust across the BLR and NLR. The H$\beta$/H$\alpha$ ratio
was, within the uncertainties, constant in time.

If dust extinction were the cause of the disappearance of the broad
emission lines, the fading of the continuum and the
disappearance of the broad emission lines would be simultaneous 
events. However, Penston \& Perez (1984) observed that the decrease in emission
line widths in NGC 4151 and 3C 390.3 {\em followed} the fading of the continuum. 
Furthermore, Inda et al. (1993) argue that the X-ray variability 
in 3C 390.3 is unlikely to be caused by changes in the nuclear obscuration 
by cold matter because the tenfold intensity reduction observed by EXOSAT 
was not accompanied by marked changes in the spectrum (eg. a deep Fe K-edge). 

Also, if the difference between Seyfert 1s and
Seyfert 2s were the result of angle dependent dust extinction, 
then one would expect a continuous change in the 
continuum and emission properties from Seyfert 1s through the intermediate
Seyferts to Seyfert 2s contrary to what is observed.
Ulvestad (1986)
studied 11 Seyfert 1.8 and 1.9 galaxies as well as Seyfert 1.2 and 1.5
galaxies. He found that the ratio of radio to featureless optical
continuum luminosity in type 1.2, 1.5, 1.8 and 1.9 galaxies lies between
the low value found for Seyfert 1s and the much higher radio/optical ratio
for Seyfert 2s. However, there is no significant difference in this ratio
between the type 1.2/1.5 nuclei and the type 1.8/1.9 nuclei.
He further claims that it is unlikely that this trend is caused
by increasing dust obscuration of the optical continua  along the 
sequence from Seyfert 1 through to Seyfert 2 galaxies because the featureless
optical continua of the intermediate Seyferts  and type 2 Seyfert
galaxies do not appear to be highly reddened.  
   
These observations indicate that profile changes must be caused by intrinsic
modifications in the source. This in turn implies that the BLR is very close to
the nucleus and that the emission line region that contributes most to the 
profile is not large in extent. 
{\em Our model BLR of NGC 4151 fits these criteria 
exactly.}
The changes can be explained by the suppression of cloud formation at low
luminosity which we predict. For a cloud density of $3\times
10^{12}r_{16}^{-2}$ cm$^{-3}$ and constant cloud column of $10^{23}$ cm$^{-2}$
the BLR cloud
lifetime is shorter than the light travel time for $r <  10^{17}$ cm. 
Thus the light travel time determines the transition timescale for
intermediate Seyferts.
Temporary changes in the classification of NGC 5548 and NGC 1566
can also be explained
by the destruction (or non-creation) of broad-line clouds. 
A test would be to look for a reduction in the
scattered polarised  flux in the broad lines at low luminosity. 

Note that we should distinguish here between the suppression of cloud formation
in our model of NGC 4151 in its low luminosity state, which is a consequence of
the substantial variability of the source, and a general suppression of 
cloud formation in low luminosity systems, which in our model occurs only for
the ILR clouds. 
Thus, according to the model, large variations in accretion
rate could suppress cloud formation for up to the mass inflow timescale from the
BLR, but at steady low luminosities the BLR clouds should always be present.
As a consequence we predict the existence of true Seyfert 2 types in 
variable systems.

Support for the `missing BLR' scenario comes 
from the observed typical lack of polarisation in the featureless continua 
of some Seyfert 2 galaxies, because if a hidden region
were to produce the continuum, it could only be observed by scattering,
and should therefore be significantly polarised. 
The detection of narrow line radio-loud or 
radio-quiet objects which do not show a large column density in their X-ray 
absorption and possessing a variable, unpolarised continuum should settle  
this point.      

At present there is observational evidence for few such objects. Kay (1994)
carried out a near UV spectropolarimetry survey of 50  Seyfert 2 galaxies and 
failed to detect in most objects a significant amount of polarisation in the 
featureless continuum. Only a few objects in her sample could possibly posses 
a hidden BLR. Tran (1995)
also points out the possibility that there `exist the genuinely pure' 
Seyfert 2 galaxies that do not contain BLRs and the appearance of which is 
independent of the observer's line of sight.

NGC 5506 is generally considered to be a bright Seyfert 2 galaxy and is a 
possible candidate for a true Seyfert 2. Since the source is variable, 
a weak BLR may be present at some epochs, so this classification is 
compatible with detection of, for example, weak broad H$\alpha$ in 
isolated observations (Shuder, 1980). Broad Paschen $\beta$ 
(1500 km s$^{-1}$ FWHM) was observed by Blanco et al (1990). 
Since the X-ray luminosity is high and activity 
is significant on a timescale of about $10^{3}$  seconds (McHardy \& Czerny
1987) the X-ray
emitting region is unlikely to be covered by a thick dusty torus and one can
reasonably assume that this region is observed directly as in Seyfert 1
galaxies. 
Pounds et al (1989) show that the spectrum may be fitted
with an iron fluorescence line at 6.4 keV and an iron K absorption edge at 8.3
keV. This has been taken to indicate ionised gas in the line of sight. 
Also, if the large dip
that appears between $\sim 8$ and 12 keV is due to Fe K-shell absorption, then
it corresponds to a large column density of (0.7 - 1.9)$\times
10^{23}$cm$^{-2}$ as in other Seyfert 2s. However, good fits were obtained without
an ionised absorber from enhanced reflection ($1.5 \times$), corresponding to a
`saucer-shaped' disc, and a $2\times$ solar iron abundance. The underlying
power-law slope in the reflection model is 0.9. 
Smith and Done (1995) also prefer a face-on disc for NGC 5506 and associate the
observed absorption with the line-of-sight column through the host galaxy
rather than a molecular torus.  

Pogge (1989) found that 
maps of the [OIII]$\lambda$5007 / (H$\alpha$ + [NII]) emission ratio,
which emphasises regions of high ionisation gas, reveal distinctly conical
high-ionisation regions, with the nucleus at the apex in four 
Seyfert 2 galaxies. Three out of nine Seyfert 1 galaxies 
showed evidence for spatially
extended ionised gas regions, compared with 8 out of 11 Seyfert 2s.
Pogge
argues that these results are inconsistent with the SPM.
Six Seyfert 1s and 3 Seyfert 2s are compact, with no suggestion 
of marginally resolved nuclear substructure.
These compact Seyfert 2s are at least consistent with the conclusion 
that not all Seyfert 2s are viewed edge-on and consequently that the 
BLR may be temporarily absent in some Seyfert 2s, not just obscured.

One of the difficulties associated with finding true narrow line objects 
is the uncertainty whether                                    
the absence of broad-lines is caused by the absence of
electron scatterers or by the absence of broad-line clouds. In our model, of
course, the absence of scatterers arises from the absence of clouds, so a
Seyfert without broad lines in its polarised spectrum must be a true Seyfert 2.

The conical geometry of the NLR in NGC 4151 clearly indicates anisotropy 
probably due to some obscuring material. If our model of NGC 4151 is to be 
extended to all 
Seyferts there must also be a torus of material that obscures the direct 
view of the nucleus 
at low latitudes. We turn to this in the next section.     

\section{Obscured Seyfert 1s and Dusty Molecular Tori}

NGC 4151 is an unobscured Seyfert 1 galaxy. According to the SPM unification
it must therefore be 
observed along the opening cone of the torus, that is within 30$^{\circ}$ of
face-on. 
Yet             
Merlin/VLA radio studies of NGC 4151 show that collimated ejection is
taking place along PA $\sim 77^{\circ}$ and 257$^{\circ}$ giving rise to  
a two-sided radio jet. Also, the X-ray column in our line of sight to 
NGC 4151 is large and most of the X-ray source                         
is quite obscured from our point of view. These X-ray and radio properties
are more suggestive of an edge-on Seyfert than a
face-on one.

If we are to include NGC 4151 in a unification scheme therefore, we must 
adjust the
parameters of the obscuring material. We begin by presenting some further
evidence in favour of such adjustments by showing that a dusty molecular torus
with small fixed opening angle cannot be a universal feature of AGN, even
excepting NGC 4151.\\
 \\
\noindent  (i) X-rays: Lawrence (1991) has argued that much of the X-ray
absorbing material in AGN is free of dust, and that narrow-lined and reddened
broad-lined objects occur more frequently at lower source powers. There must be
a large range of geometrical thicknesses and optical depths in the absorbing
material. He speculates that the obscuring material is not a molecular
torus but an expelled shell of gas. A luminosity dependant flaring of the
obscuring disc could also explain why
the fraction of narrow-lined AGN is a decreasing function
of source power. Reichert et al. (1985) find that when
comparing the X-ray column densities with optical reddening indicators
little, if any, dust can be associated with the X-ray
absorbing regions. Mulchaey et al. (1992) attribute the observed range in
column densities in a sample of Seyfert 2s to intrinsic variations in the 
thickness of the obscuring
medium and a variation of the angle to the line of sight. 
They also state that if all Seyferts have a 30$^{\circ}$ opening angle,
as required by the standard model, and if all tori are transparent to hard
X-rays, then the Seyfert 2s would overproduce the hard X-ray background. A
30$^{\circ}$ angle also conflicts with the observed low polarisation of the
featureless continuum, which suggests a larger opening angle 
(Antonucci  1993).\\

\noindent (ii) Radio galaxies: A range of torus opening angles 
(at the very least)
also seems to be indicated for FRII type
radio galaxies and quasars (Heckman et al. 1986, Hutchings 1987). \\
\\
\noindent (iii) Multi-band spectra: The lack of diversity in the shape of the 
optical-ultraviolet continuum 
of Seyfert 1 galaxies, radio-quiet and radio-loud quasars argues against 
large amounts of dust extinction (Neugebauer et al. 1984, 1986), 
since the 
effect on the continuum slope is likely to be large, even for a small amount 
of dust. 
Baker \& Hunstead (1995) have presented low resolution, composite 
optical spectra for
60 (radio-loud) quasars. They found that with increasing inferred 
viewing angle to the radio-jet axis, the optical continuum steepens, 
the 3000\AA  $\;$ broad emission feature decreases in relative strength, 
and the narrow-line equivalent widths and Balmer decrement increase. 
They also found that reddening is considerable ($A_{v} \sim 2.4$) in most  
lobe-dominated quasars. A significant amount of dust must therefore 
lie within the opening angle of the torus and there may be no well-defined 
opening angle at all. If radio-quiet quasars (QSOs) possess as much dust as 
the radio-loud ones and have gone undetected at other wavelengths then up to 
80 \% of radio-quiet quasars could be obscured by dust up to $A_{B}=5$ mag. 
(Webster et al. 1995).
This would be difficult to reconcile with our model. However, 
Boyle \& di Matleo (1995) use a sample of X-ray selected QSOs to place 
a limit on the intrinsic amount of dust of $A_{B} < 2$mag.  
They also observe a similar spread in the X-ray-to-optical flux ratios in 
an optically selected sample of QSOs. Their interpretation is that most of 
the scatter is probably due to effects other than dust obscuration. 
Hence, it seems unlikely that a significant population of highly obscured 
QSOs exist in X-ray selected samples. These studies can be reconciled if 
radio-loud quasars have systematically much larger amounts of associated 
dust than radio-quiet QSOs. 

The same argument applies to the continua of Seyfert 2 nuclei. 
Their non-stellar optical, ultraviolet and X-ray continua 
are very weak, but their ratios are not too different from those in Seyfert 1s.
Therefore, the overall continuum shape is similar in both groups.
There is also no evidence for dust as a major constituent in terms of 
severe modification of the intrinsic spectrum in blazars 
(Sanders et al. 1989).\\
\\
\noindent (iv) Infrared: However the near IR distribution in most 
AGN show features that can be
attributed to hot dust emission: a minimum at about 1 $\mu$m (Sanders et al.
1989) and a bump peaking at 3-5 $\mu$m (Edelson \& Malkan 1986, Robson et al.
1986). According to Giuricin et al. (1995) the lack of a good correlation
between the near and mid IR on the one hand and the 100$\mu$m emission 
on the other is interpreted to mean that the bulk of the FIR band comes 
from the host galaxy while the NIR and MIR is nuclear.  
NGC 1068 is considered to be 
exceptional in a number of ways. The far infrared is resolved
and originates from a 3 kpc diameter star formation ring (Telesco et al 1984; 
see also Rodriguez Espinosa  et al.1986,1987, Clement et al 1988 for 
similar conclusions in other systems).
The X-ray column is high,
$> 10^{24}$ cm$^{-2}$ (Mulchaey et al. 1992) possibly as a consequence of the
edge-on view through the obscuring ring.
The origin of the mid-infrared is still being debated, but is generally thought
to be associated with dust emission in both radio loud systems (Heckman et al.
1992,1994, Efstathiou \& Rowan-Robinson 1995) and Seyfert galaxies 
(Giurian et al. 1995, Maiolino et al. 1995, Heckman et al.1995).
Some variability timescales are also in agreement with dust emission 
(Bregman 1991). \\
\\
How can we extend our picture of NGC 4151 to be compatible with this evidence?
Viewed  from close to the plane of the disc the line of sight to the
BLR and ILR passes through the flared disc. The optical depth through the
disc is determined by the density at the inner edge.
The density of hot material near the
disc surface is of order $6 \times 10^{8}r_{16}^{-2}$ cm$^{-3}$ 
cm, ignoring parameters of order unity 
(Cassidy \& Raine 1996). 
If  the obscuring disc consists of a hot isothermal externally -heated atmosphere
above a standard $\alpha$-disc the density falls off
along a line of sight through the atmosphere as $r^{-15/8}e^{-{\rm const}/r}$, 
which is steeper than $1/r$. 

The column to the inner BLR at $10^{16}$
cm is then no  more than $\sim 6 \times 10^{24}$ cm$^{-2}$. This would be 
sufficient to account for the
reduction in hard X-ray flux by electron scattering out of the line of sight in
the majority of obscured systems. 
However, it is clear that in order that it be completely suppressed, 
the direct soft X-ray emission requires photoelectric absorption,
hence cooler gas. In addition, 
the broad-line emission is completely blocked
from direct view by 10 - 100 magnitudes of extinction.  It is therefore
impossible to contrive a model that does not involve dust. The above
calculation shows with the normal gas/dust ratio there would be 
sufficient material
in the line of sight in the disc to produce the required extinction (a column
of 6$\times 10^{24}$ cm$^{-2}$ corresponds to an $A_{v}$ of 500), but in the
externally illuminated disc atmosphere it is
too hot for the dust to survive. The simple picture of a dusty, but otherwise
standard, accretion disc extending inwards into the BLR region
is untenable. Nevertheless
we have in NGC 4151 direct evidence for the presence of dust not in the line of
sight through observations of Si and forbidden Fe emission lines in the
infrared (Thompson 1995). Thus, we may assume that there is some
cold dusty material that would be in the line of sight at low inclinations only. 
This material
must be between the BLR and NLR, beyond the region of disc flaring associated
with the BLR (so could be in the ILR region).
The response of this material to its environment will be important when we
consider highly luminous Seyferts and quasars and is
discussed in section 6.   
                                 
\section{An alien view of NGC 4151} 

We can now consider how
the model of NGC 4151 would appear viewed first from angles more face on to 
the disc, and then from closer to the disc plane. We shall argue that
this is the basic structure of all AGN. Referring to figure 1 we see at edge-on
angles to the line of sight the direct view to the nucleus is blocked. Such a
system would be classified as a Seyfert 2 with scattering into the line of
sight giving polarised broad-line component and a weak X-ray continuum.
At larger angles to the disc the line of sight passes through the BLR clouds
giving objects such as NGC 4151 and NGC 3516. At somewhat larger inclinations
the X-ray source is covered by the warm absorber only: an example is NGC 5548.
Face on objects have no absorption and relatively narrow lines. 
 
\subsection{Line Profiles and Transfer Functions} 
The BLR is usually assumed to be a spherical distribution of clouds. To
obtain a sufficiently narrow transfer function for the response of the broad
lines to changes in the ionising continuum in NGC 5548 Ferland et al.(1992)
has proposed that 
anisotropic emission from the clouds must be considered. In NGC 4151 the
transfer function is narrow (Netzer 1990) but, in contrast to NGC 5548 
the response is
immediate. To match this behaviour it would be necessary to assume the emission
from clouds in NGC 4151 is not anisotropic. Indeed, with anisotropic 
emission it would be 
difficult
to see how the MgII and CIV profiles in NGC 4151 could be similar 
(except in the
extreme wings). An alternative possibility is to assume a flattened cloud 
distribution
for {\em both} cases. This gives a             
narrow transfer function for short continuum pulses, but different delays for
different lines of sight. This can
be seen by comparing the non-zero delay from a face-on disc with 
the immediate response from a sphere of                                            
the same radius. A narrow width and immediate response are confirmed by 
numerical computation for the case of NGC
4151 (Cassidy \& Raine 1996). Note also that the we obtain different lags for
different continuum pulse lengths which again is a consequence of a
non-spherical geometry. 
For an edge-on line of sight a flattened configuration exhibits an immediate
line response to continuum changes. More face on systems show only a
delayed response as in NGC 5548 (Ferland et al 1992). 
In figure 3 we show a 
comparison of the transfer
functions for a model system with an inner disc radius of 
4$\times 10^{16}$ cm
and a disc inclined at  24$^{\circ}$ to the line
of sight with one at 75$^{\circ}$. At the higher inclination 
(75$^{\circ}$) the lag of the initial line response to the continuum  
is about one
week, but note that the peak in this case comes earlier than that for 
lower inclination. 

Both Done and Krolik (1996) (for NGC 5548) and Ulrich and Horne (1996) 
(for NGC 4151) 
find some evidence that the peak of the transfer function for the red wing  
comes earlier than that for the blue wing of the line. 
The existence of a
reflected component to the lines (as a result of reflection in the disc)    
explains the different responses of the red
and blue wings with that for the red wing being more disc-like (narrow and
sharply peaked). The transfer function for the blue wing is broader. These
results arise from the non-spherical cloud distribution.
For a spherical
geometry, a contrived velocity field is required to obtain this behaviour
(eg Mathews 1993).
                                                                                
Note that although the line
profile widths decrease for more face-on discs, they do not become exceptionally 
narrow when viewed face-on because the clouds have a velocity component normal
to the flared disc. In fact, the narrowest lines occur for a  
disc at $\sim 75^{\circ}$ as a result of flaring. 
We note that, if the disc surface is smooth, the reflected component 
will tend to be polarised in the plane of the disc. However, by construction, 
the disc surface in the BLR is far from smooth and this will tend to 
de-polarise the reflected component.

\subsection{Predictions for X-ray Absorption}                 
In section 2.2 we considered the X-ray properties of the model
specific to NGC 4151.
We now extend the discussion to different lines of sight.
Over a range of angles, the line of sight to the X-ray source
will intersect the outer region                      
of the disc and the broad-line clouds. At larger angles, it will pass through 
the ILR clouds and the warm debris from 
clouds as they are disrupted and heated to the Compton temperature.
We would expect
these systems to exhibit complex X-ray absorption. A few systems like NGC 4151
should contain the cold outer screen of one or both of the flared disc and ILR
clouds, either  totally or partially
covering the central X-ray source. In these systems there will be a combination
of partial covering by the BLR clouds, which should vary with the BLR lines,
and by a warm absorber. At larger angles to the disc this complex system should
be replaced by a warm absorber or, possibly, a combination of warm 
absorber and ILR.
Since the BLR is directly                          
visible, these will be nuclei of intermediate Seyfert type (NGC 5548
is a typical example) or, at higher luminosity, Seyfert 1s. 
Note that the cloud debris exists over a range of temperature so
detailed modelling is required to confirm that this material can be both the
source of the electron scatterers (which requires a temperature approximately
$3 \times 10^{5}$ K) (Anntonucci, 1993)) and the warm absorber (which needs 
$T \geq 10^{5}$K (Pounds et al. 1990)).

To compare these predictions with observations we note the following points.
The X-ray continuum in radio-quiet systems is usually fitted by an underlying 
power-law with an average photon number index
of 1.8 - 1.9 once a number of features have been removed. These features
include an iron
emission line near 6.4 keV, a `hump' above $\sim 10$ keV and in some Seyferts a
`warm' iron K-absorption edge (Nandra \& Pounds 1994). 

The K-absorption edge at 8-9 keV indicates a substantial column density of 
highly ionised matter, a so-called `warm absorber', in the line of sight to 
several AGN, including some with otherwise `bare' Seyfert 1 nuclei (Pounds 
et al. 1990, Smith \& Done 1995). 
This warm absorber can also account for the spectral variability
and soft excess evident in Ginga and some EXOSAT data.
The warm absorber can change rapidly only
if its density is high, $\geq 10^{7}$ cm$^{-3}$ (Pounds et al. 1990).
standard spectrum).
                                                                
Nandra and Pounds (1994)
find that their lowest luminosity warm absorber (NGC 4051) is at a maximum
distance of $\sim 10^{17}$ cm from the central engine and for the highest 
luminosity warm absorber (MCG-2-58-22) they obtain
an upper limit of $\sim 10^{20}$ cm with a covering factor of the order
$\geq$ 50 per cent. This places the warm absorber at a similar distance from
the nucleus as the broad-line region in the disc-wind model. ASCA observations
of variable absorption in MCG-6-30-15 have been used to give a
(model-dependent) location for the warm absorber in or just beyond the BLR
(Reynolds \& Fabian 1995). If the warm absorber is not present
then the native spectrum will exhibit a large soft excess (relative to the

At even higher latitudes we expect to see X-ray absorption 
associated only with the NLR scatterers (if these exist) and with the galaxy. 
Such objects will show relatively narrow broad lines. They may also possess
an intrinsic {\em ultra-soft} spectrum particularly if, as is the case 
in many of the early EXOSAT observations, the power law slope is 
extrapolated from the hard X-ray band. ( And correspondingly the hard tail 
is missing in such sources.)
It is
interesting to note that Boller et al. (1996) find narrow line Seyfert 1s have
generally steeper soft X-ray slopes with rapid soft X-ray variability and
therefore significant electron-scattering of the X-rays seems unlikely in these
systems. There is in addition no evidence for large neutral H column densities
in excess of the galactic column. Conversely,
soft X-ray selected samples of Seyfert 1s contain relatively large numbers of
narrower line Seyferts (Stephens 1989, Puchnarewicz et al 1992). 

\subsection{Scattering into the line of sight}  
The optical to soft X-ray continuum in Seyfert 2 galaxies is partly radiation
scattered into the line of sight. The opening
angle of the cone of light incident on the scattering region 
is deduced to be $\leq 30^{\circ}$ (Antonucci, 1993)
from the fact that the scattered light is polarised 
perpendicular to the radio jet axis. (The direction of polarisation defines the
E-vector prior to scattering.)
But this does not require radiation to be {\em blocked} within a solid cone of
this angle; the result can also be 
achieved if the {\em scatterers} are associated with a restricted conical
shell. In  
our model the BLR debris provides just such a conical shell. This leaves the
opening angle of the obscuring torus as a free parameter. For consistency
with our model of NGC 4151 the half angle of the cone unobscured by dust 
must be at least 60$^{o}$ (since the disc flares to about 30$^{\circ}$). 
Therefore, at the luminosity of NGC 4151, obscured Seyfert 1s make up about
half of the observed systems rather than the 80\%
of the standard model. Since it is in this luminosity range that Seyferts make
the most important contribution to the X-ray background this eliminates the
problem of overproduction of that background associated with a constant 
small opening
angle (Mulchaey et al. 1992).

Light within the inner unobscured cone escapes to the NLR giving a conical
illumination to the NLR clouds (Tadhunter \& Tsvetanov 1989, Wilson et al.
1988).
There is evidence in NGC 1068 for scattering of this light 
from material at $\sim 100$ pc (Antonucci 1993)  
 - as well as scattering from closer in: the
multiplicity of the Fe K lines requires the existence of a hot $T \sim 4 \times
10^{6}$K component of the scattering medium in addition to the warm $T\sim 2
\times 10^{5}$K scattering cloud required by the polarised light. The relative
strengths of the Fe K lines indicate that the two components scatter comparable
amounts of radiation (within a factor of a few) (Marshall et al. 1993). In
their preferred model the warm component extends from 20pc to 100pc while the
hot component begins at 1-2 pc.

\subsection{The Intermediate Line Region}
An intermediate line region (ILR) is essential in our picture of NGC 4151. 
If this is
to be a typical galaxy an ILR must be present in all comparable systems 
(although it need
not be a distinct region: the BLR and ILR may be continuous). There is a growing
body of evidence to support this view.\\
\\
\noindent (i) Individual objects:\\ 
(a)3C390.3 is a broad-line radio galaxy exhibiting  a Seyfert 1 type 
spectrum and,
like NGC 4151, has appeared close to a Seyfert 2 (really, of course, in
this case a NLRG), for example during 1979-80 (Barr et al.1983).
Clavel \& Wamsteker (1987)  argue that the BLR 
gas cannot be responsible for the relatively narrow emission lines 
in the low state,
which must come form an intermediate region.  Zheng (1996) 
found that the narrow component (FWHM $\sim 2000$ km s$^{-1}$ in the 
Ly$\alpha$ and CIV lines varied by a factor of approximately 2.5 
with the CIV/Ly$\alpha$ flux ratio of 0.4, whereas for the broad 
component this ratio is around unity.\\
(b)Indications of the existence of an additional emission line region 
at greater
distances than the BLR were also found in 3C382 (Yee \& Oke 1981),
3C120 (Oke, Redhead \& Sargent 1980), Fairall 9 (Stirpe et al. 1989), 3C 445
(Crenshaw, Peterson \& Wagner 1988), NGC 7469 (Bonatto et al. 1990),  
and NGC 5548 (Wamsteker et al. 1990). \\
(c) Filippenko (1985) investigated Pictor A, a radio galaxy with                   
features of a LINER, PKS 1718-649, a classical LINER, and the QSO
MR 2251-1878 and found that  in all three nuclei 
the great strength of [OIII] $\lambda
4363$ relative to [OIII] $\lambda 5007$ implies $T_{e}$ $>$ 50,000 K,
which is incompatible with photoionised low-density clouds ($n_{e} \sim
10^{2} - 10^{4}$ cm$^{-3}$).
He suggested that this dilemma vanishes 
if relatively dense clouds ($n_{e} \simeq 10^{6} - 10^{7}$ cm$^{-3}$) 
exist in the narrow line regions.(See also Filippenko \& Halpern 1984.)  
We suggest that in our model the high density contribution to
the [OIII] line comes from the ILR while the NLR densities are similar in all
AGN.\\

\noindent (ii) Surveys: \\
(a) Francis et al (1992) applied principal component analysis to a QSO 
sample taken from the Large Bright QSO survey and found that three 
principal components account for approximately 75\% of the intrinsic 
variance in the sample. The analysis supports a two-zone line emitting 
model with a BLR and an additional region producing the line core, 
FWHM $\sim$ 2500 km s$^{-1}$. They point out that this line core region 
is not the traditional NLR.\\
(b) Van Gr\"{o}ningen and de Bruyn (1989) found broad wings in 
[OIII]$\lambda$5007 in 
10 out of 12 objects in their sample of Seyfert 1 galaxies. 
From the ratio to [OIII]$\lambda$4363 and H$\beta$ in 4 of the 10 they 
deduced the existence of a transition zone
intermediate between the BLR and NLR with a mean density of 
5$\times 10^{6}$ cm$^{-3}$ at a minimum radius of 0.8 pc.  \\
(c) Wills et al. (1993) investigated 123 high luminosity AGN. 
They found that the CIV profile consists of a core with characteristic 
width of approximately 2000 km s$^{-1}$ FWHM plus a broad base component. 
The core component is attributed to an ILR.\\
(d)In those cases where they could measure strong wings Brotherton et al. 
(1995) used principal
component analysis on the H$\beta$ profiles of 41 radio-loud quasars and found
that the second most significant variation in this emission line appears to
involve an intermediate width ($\sim 2000$ km s$^{-1}$ ) component with a small
redshift. The strength of this component is correlated with the strength of
[OIII] $\lambda 5007$.\\ 
\\
(iii) The narrow lines:\\
Narrow emission line asymmetry is observed in most Seyfert 1 and Seyfert 2
galaxies. The narrow line profiles in these objects are usually smooth with 
a clear tendency for
the blue wing of some lines to be stronger. This blue asymmetry is 
most noticeable in lines of higher excitation or critical de-excitation
density. Also, the higher excitation lines are usually found to be broader
and more blueshifted than the low excitation lines (Whittle 1985, 1988).  
In our model such asymmetries may be explained, at least in general 
qualitative terms, by contributions from the ILR
to the narrow emission lines.

This ILR contribution
to the narrow lines might also be the reason for differences observed in
the spread of line profile shapes in a given object, $\Sigma$, in a sample
of Seyfert galaxies. Whittle (1988) found that Seyfert 1s have high $\Sigma$, 
Seyfert 2s low $\Sigma$ while intermediate Seyferts have values in-between.
Whittle (1985) suggested that galaxy or
bulge mass may be the important underlying variable but in the later paper 
(1988) could find no correlations between line profile variations and host
galaxy properties. The
observed trend was accounted for in a two component picture of
the NLR (Mobasher \& Raine
1989), similar to the NLR/ILR picture considered here in as much as the second
component provides the higher velocity, higher density material. 

Whittle also found a lack of correlation between [OIII] line width 
and the nuclear nonthermal luminosity, the strength of the BLR or the 
ionisation and physical state of the NLR gas.
For the Seyfert 1 sample, however, correlations were found between 
broad (H$\alpha$ and H$\beta$) and narrow ([OIII]) line widths.
The correlation with [OIII] is stronger with the 
broad-line core widths than with the broad-line wings. Similarly, the broad
line widths are more
strongly correlated with the narrow line base widths than the narrow line 
core widths (Whittle 1985; Wilson \& Heckman 1985).    
In our model a significant contribution to the broad-line core emission 
comes from the ILR as do the base widths of the higher critical density 
narrow lines (particularly [OIII]).  
So we expect a correlation between the narrow line widths and the
broad-line core widths and between narrow line base widths and broad-line widths.
The ILR emission provides a possible basis for a
simple explanation of these observations and the lack of correlation
between line profile differences and host galaxy properties. Absence (or
obscuration) of the
ILR in NLRGs might also provide an explanation for the mismatch of the [OIII]
$\lambda 5007 $ emission in BLRGs and NLRGs, while the [OII] $\lambda 3727$
luminosities (which come only from the NLR) in these galaxies match each 
other well (Hes et al. 1993).

\subsubsection{Coronal Lines}
Coronal lines are forbidden transitions within the ground terms 
of highly ionised species ($h\nu_{ion} \gg 100$eV) (Oliva et al. 1994) 
which can be formed either by a hard UV photoionising continuum 
(Grandi 1978, Korista \& Ferland 1989) or in a hot ($T \sim 10^{6}$K) 
collisionally ionised plasma (Oke \& Sargent 1968, 
Nussbaumer \& Osterbrock 1970). A combination of shocks and 
photoionisation has also been suggested (Viegas-Aldrovandi \& Contini 1989). 

The [Si VI]  $\lambda$ 1.962 line seems to be associated with 
the active nucleus as this line is found in Seyferts but not in 
starburst galaxies (Marconi et al 1994).
Spinoglio \& Malkan (1992) placed the coronal line region just 
outside the BLR. Thompson (1995) considers that in NGC 4151 the 
coronal lines arise near the interface between the dusty torus and 
the BLR mainly on account of the large line widths.
The critical density for the [Si VI] line at 1962 $\mu$m is 
$\sim 10^{8}$ cm$^{-3}$ (Greenhouse et al. 1993, Oliva \& Moorwood 1990). 
In our model these densities are associated with the ILR, 
the dispersing BLR clouds and shocked material in the BLR. 
Since  the [Si VI] line widths span a large range (Giannuzzo et al 1995), 
from very narrow profiles (FWHM $\sim$ 300 km s$^{-1}$) to widths 
approaching those of the broad lines, the main contribution to the 
[Si VI] line may  come mainly from different regions in different galaxies.

\section{A Unified AGN Model}
                                                                  
We have so far seen that viewed at various angles our picture of  NGC 4151
gives rise to  obscured Seyfert 1s 
with a range of obscuring columns, a class
of intermediate Seyferts and Seyfert 1s with varying degrees of partial
covering and absorption of the X-ray source and unobscured Seyfert galaxies.
We now wish to apply the model to objects at lower and higher luminosities than
those attained in the various states of NGC 4151.
We therefore need to know how the
intrinsic structure varies with luminosity. There are at least two
predictions relevant to a unified model. These are that
both the disc flaring and the injection of cloud material from the disc
are functions of luminosity. For the latter we can adapt the discussion of 
Smith and Raine (1985, 1988, following Begelman et al. 1983): external
illumination drives a thermal disc wind which here fuels the ILR (not the BLR
as in our earlier papers); for an incident luminosity        
$L/L_{Edd} < 0.03T_{8}^{1/2}$,                                             
where $T_{8}$ is the Compton temperature of the incident radiation in units of
$10^{8}$K, a strong                                                       
disc wind does not form and we expect there to be little or no ILR gas.
We have already seen that the BLR clouds are not present at low luminosity in a
source with time varying luminosity.     

Figure 4 shows our extension of the unified picture beyond the range of
luminosities appropriate to NGC 4151. We summarise it here and give more details
in the following sections. It is reasonable to assume that a steady high 
luminosity is 
associated with high wind power. Then we predict that the disc in the BLR 
does not flare at high luminosity (section 6.1). 
Thus there are no NGC 4151-like absorption    
systems in high luminosity objects, but we get instead, at the right viewing
angle, the broad absoption line quasars (BAL QSOs). At low steady 
luminosity the BLR survives, but the ILR does
not. We also expect (although we do not show here) that the absence of a flared
disc
exposes the dust region to the full force of the nuclear wind, giving rise 
to strong
shocks (in the wind and the dusty material) and strong heating.  Therefore 
the dust obscuration is less important at high luminosity. The fraction of
obscured systems should therefore decrease with increasing continuum power. The
model is consistent with a luminosity sequence: LINERS, True Seyfert 2s, 
Intermediate Seyferts, Seyfert 1s, QSOs. 

Note that the similar upturns in the near infrared that is seen in 
almost all quasars and Seyferts is evidence for similar covering 
factors of dust in these systems. This is supported by the fairly 
small range in the relative sizes of the blue bump and the infrared bump  
found by Sanders et al. (1989) in 109 bright quasars in the PG survey. 
This is compatible with our picture, because the flux falling on the 
disc is only weakly dependent on its shape: a disc flattened by a 
significant wind will experience less direct irradiation but a larger 
component due to scattering in the wind. We find an effective covering 
factor that is therefore always in the region of 10 \%.

\subsection{Luminosity Dependence of the Model}

We compute first how the disc flaring depends on luminosity assuming a steady
state. Let the angle between the radius vector to a point on the disc surface
and the tangent vector at that point be $\alpha$. If the ram pressure in the
radially flowing wind is $P_{m}$ then the interaction with the disc destroys 
momentum normal to the disc surface at a rate $P_{m}\sin \alpha$ per unit area.
But the average pressure on the disc surface is $fP_{m}$ where $f\sim 3 \times
10^{-3} (L/L_{m})$, omitting factors of order unity, $ \sim
10^{-4}$ (Cassidy
and Raine 1996). Thus $f=\sin \alpha$ where, if $z=z(r)$ is the disc surface
as a function of radius $r$ {\em in the disc},
\[ \alpha = \tan^{-1}(dz/dr) - \tan^{-1}(z/r) .\]
Taking the small angle approximation therefore gives for the disc surface 
\[ \frac{dz}{dr} - \frac{z}{r} =f \] 
or
\[ z=(h_{0}/r_{0} + f \log (r/r_{0}))r  \]
where $h_{0}$ is the disc height at some reference point $r_{0}$. In this
region the disc height increases close to  linearly with radius. 

As we move out further in the BLR region, beyond a radius $r_{*}$ (say),  
the momentum extracted from the wind
to accelerate the clouds cannot be neglected. The loss of momentum in an
element of wind as it entrains clouds at the disc surface between $r$ and
$r+dr$ is proportional to
the mass injection rate in clouds, $d\dot{M}/dr$, and inversely proportional to
the residence time in the element, $v_{n}/\dot{v}.$ Since (see Cassidy \& Raine
1996) $d\dot{M}/dr \propto 1/r$ and the cloud acceleration, $\dot{v} \propto
1/r^{2}$ we have, in this region, $r>r_{*}$, that the actual ram pressure of the 
wind, $P_{m}^{'}$, is 
\[ P_{m}^{'} \propto r^{-3} = P_{m}r_{*}/r .\] 
Thus, the disc pressure, which by definition is $fP_{m}$, must now balance
$\sin \alpha P'_{m}$, from which we derive, as above,
\[z=r[h_{BLR}/r_{BLR} + f(r-r_{BLR})/r_{*}],\]
where $h_{BLR}$ is the height of the disc at the outer edge of the BLR. We take
this as a scale height at $r_{BLR}$. 
Therefore, in this region the disc rises more strongly into the wind in order to
balance the pressure of heated material at the surface and the surface flares
until it reaches a scale height at the outer radius of the BLR. Clearly the
radius, $r_{*}$, at which the disc begins to flare will be larger, for a given
mass loss rate in clouds, the larger the momentum in the wind. 
Provided the momentum flux in the wind increases with the luminosity of the
source more strongly than linearly (which we know to be the case in NGC 4151
from the line profile variations) then $r_{*}$ will increase with luminosity
until at high luminosity the disc does not flare at all and the BLR merges
directly into the ILR. We see this as follows.

At the transition radius $r_{*}$ we have, neglecting logarithmic terms,
\[h_{*} = r_{*}[h_{BLR}/r_{BLR} +f(r_{*}-r_{BLR})/r_{*}] \sim h_{0}r_{*}/r_{0}.
\]
Solving for $r_{*}$ we get
\begin{eqnarray*}
r_{*} & = & r_{BLR}[1-(h_{0}/r_{0}-h_{BLR}/r_{BLR})(1/f)]^{-1}\\
      & \sim & r_{BLR}^{2}f/h_{BLR}.
\end{eqnarray*}
We have $h_{BLR} \propto r_{BLR}^{3/2} $ for an isothermal disc, and $r_{BLR}
\propto (L/L_{m})$, ignoring logarithmic terms (Cassidy and Raine 1996). Thus,
finally, 
\[r_{*} \propto (L/L_{m})^{3/2}, \]
and the flared region is smaller for higher luminosities.

A further consideration is the importance of radiation pressure
acceleration of the clouds. For the acceleration, $\alpha_{w}$, due to the
wind, having speed $10^{10}v_{10}$ cm s$^{-1}$,  
and the acceleration due to radiation pressure for a completely absorbing
cloud, $\alpha_{r}$,  we find  the ratio
\[ \frac{\alpha_{r}}{\alpha_{w}} = 0.3 (L_{r}/L_{m}) v_{10} .\]
So, unless the energy flux in the wind is comparable with that in radiation,
radiation pressure driving is important. 
Since radiation pressure and ram pressure have the same radial
dependence this makes no difference to the line profiles, except that changes
in luminosity are now linked directly to changes in profile width, rather than
through assumptions on the relation between the momentum carried by the wind
and by the radiation field. (For NGC 4151 our profile fit in the high state gave 
$L_{r}/L_{m} = 0.2$ so radiation pressure acceleration is unimportant. It may be 
significant  in lower luminosity states since $L_{r}/L_{m}$ increases as $L_{r}$ 
decreases.)
                                      
\subsection{Lower luminosity systems}

The sub-class of LINERs that are the lowest luminosity AGN (Heckman 1986)
appear in our model 
at low values
of \.{M}.  (Note that we are concerned only with the {\em AGN-type} 
LINERS in the following. without  prejudice to other types.)
 Over 1/3 of LINERs studied in a survey of intrinsically faint Seyfert 
nuclei have detectably broad H$\alpha$ wings (FWHM a few thousand 
km s$^{-1}$), reminiscent of the broad emission lines that define 
type 1 Seyfert nuclei (Filippenko \& Sargent 1986). Heckman (1980) 
had already pointed out that many LINERs appear to form
a natural extension of Seyfert 2 galaxies to smaller values of 
[OIII]/H$\beta$. Moreover, the           
narrow emission lines in a given nucleus often have markedly different
widths as observed in the NLRs of Seyferts. For example,
[OI] $\lambda$6300 is far broader than the [NII] and [SII] lines in
the nucleus of M81. The [OI] line has $n_{crit} \sim 1.4 \times 10^{6}$
cm$^{-3}$, whereas the much narrower [SII] $\lambda \lambda$6726,6731 
lines have $n_{crit} \sim 2.1 \times 10^{3}$ cm$^{-3}$. This confirms that
the gas density in the NLR in LINERs is increasing with decreasing distance 
from the centre (Filippenko \& Halpern 1984) in the same way as in the 
NLR of Seyferts (see Whittle 1988).   
Further evidence
for a mini-Seyfert nucleus at the centre of LINERs comes from observations
of NGC 1097 by Storchi-Bergmann et al. (1993). They found that recently
broad double-peaked H$\beta$ and H$\alpha$ components appeared in the 
spectrum of the LINER.

For the lower luminosity systems our model predicts that with 
highly {\em variable} luminosities the BLR clouds
will be wholly or partially absent in low luminosity states. This is because the
inner radius of the BLR depends on the ratio of the current accretion 
power output
to the local accretion rate; if this is small then the external illumination
does not affect the disc and no BLR clouds are formed (Cassidy \& Raine 1996). 
It also predicts,
however, that in steady systems the BLR is always present, but that the ILR
will be weak or absent below about $\l_{r}= 0.03 L_{Edd}$ at which the inner and
outer radius of the thermally driven disc wind coincide (Begelman et al. 1983).   
Thus we distinguish between those Seyfert 2s that are 
really hidden type 1s and 
`genuine' type 2s. In hidden type 1s  the BLR will be
obscured and the ILR may be. In true type 2s, for the most part, we 
expect both the BLR and ILR
to be absent, but the population of these will be rare if variability decreases
with decreasing  luminosity.

Calculations by Ferland and Netzer (1983) and Halpern and Steiner 
(1983) indicate that LINER-like spectra could indeed result from 
reducing the luminosity of a Seyfert-like power-law continuum source 
by factors of 10-100 without changing the gas density or volume. We 
expect such an appearance for our steady-state unobscured 
low-luminosity systems.

\subsection{Higher luminosity systems} 
At the highest luminosities we expect a Seyfert nucleus to be classified as a
QSO. We expect the BLR to merge into the ILR, since the inner radius of the ILR
is independent of luminosity whereas the outer radius of the BLR increases
directly with luminosity (for a fixed black hole mass).
We have seen that the flared region of the disc is no longer present at high
luminosity  so it will be necessary to calculate
the effect of the nuclear wind impacting on the obscuring torus. We speculate
that this leads to a deposition of energy in the torus (e.g. by shock heating)
which will erode the dust at higher latitudes. 
Therefore we
expect obscuration to be less likely in high luminosity objects. So we
expect few narrow line QSOs (high luminosity analogues of obscured Seyfert 1s).

Instead, viewed from close to the disc, the continuum and inner clouds are
covered by the nearside outer broad-line clouds. These clouds are accelerated
to high velocity in the line of sight and so give rise to broad absorption
lines. These systems are therefore the BAL QSOs (see section 6.4). 
The BAL QSOs are the high luminosity analogues of NGC 4151 and should 
therefore show complex X-ray absorption. Because the
outer clouds are at larger radii in QSOs they have lower density and longer
sound-crossing (or survival) times. Consequently
we expect line absorption to be relatively more common in QSOs than in Seyferts.  

Since  the cloud debris is present at all luminosities, at appropriate
angles X-ray absorption by warm material should be observed.
Warm absorption has been observed in QSOs with the absorbing material being
close to the central source (Pan et al. 1990). X-radiation falling on the 
disc, either 
directly or by scattering in the wind, will still be reprocessed by dust 
at low latitudes, so we do not expect large differences in the 
reprocessed infrared.

At luminosities high enough that the BLR disc is non-flaring the width of the
CIV line is an orientation indicator (in radio-quiet objects).

\subsection{Broad Absorption Line Quasars}  
                                                                       
About 10\% of all radio-quiet QSOs                                      
possess very broad absorption features (BAL QSOs, or BALs). Usually these
are explained by broad-line clouds moving in front of each other and the source
in certain lines of sight and therefore it is generally assumed that all QSOs
possess a BAL region. The close similarity between emission line and 
continuum properties
of QSOs and BALs (Weymann et al. 1991) is also consistent with the
view that BALs do not from an intrinsically different class of QSOs. 
Broad absorption is mainly found in 
high ionisation resonance lines (eg. NV, CIV), while it is weak or
absent in MgII. The broad absorption lines possess the following relevant
properties: \\
(i) They often appear detached from the emission peak,
but usually set in by the time an outflow velocity of 5,000 to 10,000
km s$^{-1}$ is reached.\\
(ii) The absorption profiles differ widely in terms of outflow velocity,
level of ionization, velocity structure in the absorption troughs and strength
of absorption (Turnshek 1987).\\
Observations also indicate that we are looking at material which has
been accelerated radially outward (there are no red absorption troughs in
BALQSOs) and that the absorbing material is situated fairly close to the 
quasar nucleus.

An important property of BALs is that they are all radio-quiet.
Also, the optical polarisation properties in BALs appear 
to differ markedly from other radio-quiet QSOs, with BAL QSOs
having a high probability of exhibiting significant polarisation (Turnshek
1987).  

In his unified model Barthel (1989, 1991) associates the BAL regions with
the weak radio jets of radio-quiet quasars. He suggests that BALs
are viewed at small angles - the observer looks at the jet - but are
otherwise `normal' QSOs. The problem is that, with a single jet-like geometry, 
it is difficult to produce the observed symmetric NV broad emission 
lines, for example 
by resonance scattering (Turnshek 1987); i.e. the NV line should be asymmetric.

The broad profiles and the large range of different 
profile widths in BAL QSOs are not compatible with a face-on disc in our
BLR model since the emission 
lines from flat face-on discs are comparatively narrow 
with no large differences in widths between various objects. The observed 
BAL properties
are, however, compatible with edge-on discs.
At small inclinations
to the disc plane the outer BLR clouds may shield some of the inner clouds and the 
continuum source from our view. 

Note that broad absorption lines are less likely with a
flared disc (because the BLR is hidden at most angles for which the shielding
would occur), although the blue absorption feature in CIV for the special
lines of sight to NGC 4151 and NGC 3516 arises from such shielding in
our picture. 
The narrowness of the absorption dips in these objects 
results from the small BLR size and flaring of the disc. 
The smaller BLR size in Seyfert galaxies can also
explain the correlation between the absorption troughs and the 
continuum intensity in these objects, which is not normally observed in 
BAL QSOs. 


Our model predicts Seyfert-like absorption profiles (as opposed to BALs) 
for radio-loud
BLRGs, where the BLR is in the line of sight. This should apply even to the
most luminous  objects because the disc should always be flared in radio 
sources (section 7).

In our model the outer BLR clouds in QSOs are strong emitters of MgII,
whereas CIV is mainly produced in the inner part. We therefore do not
expect to see many broad MgII absorption systems. Also, we expect a
difference in the absorption velocities of CIV and MgII. Both these
predictions are confirmed by observations: MgII absorption is found in only
$\sim 15$\% of all BAL QSOs and where present it is generally less extensive in
velocity space than CIV (Woltjer 1990).
The geometry of BAL quasars predicted by our model agrees in general
with the picture proposed by
Turnshek (1987). 
                 
Voit et al. (1993) estimate that the BAL clouds have  
equivalent hydrogen
column densities  $\sim5 \times 10^{21}$ cm$^{-2}$ to $10^{23}$ cm$^{-2}$ and  
ionisation parameter, U $\sim 0.1-1.0$, where $U$ is given in terms of the
luminosity, distance and cloud density (cm$^{-3}$) 
as $U=10^{8} L_{46}/nR_{pc}^{2}$. 
These estimates are within the range of values for the outer BLR and ILR  
of the disc-wind picture. In addition, in our model of NGC       
4151 we ignored radiation pressure acceleration of the clouds. Since the
radiative driving force, like that due to ram pressure, falls off as $1/r^{2}$
the inclusion of radiation pressure makes no difference to the agreement
between the model and observation except that it makes it impossible to
determine a unique set of wind parameters. There is therefore no
conflict between our results and the conclusions of Arav et al.(1995) on the
role of radiative acceleration in the BAL region. Our picture 
for BAL QSOs is similar to those proposed recently by (Goodrich \& Miller 1995,
Cohen et al. 1995), except that we associate the BAL region with the outer BLR. 

Nevertheless there remains a significant problem for {\em all} 
theories of the BAL region, namely the apparently very high metal 
abundances compared to solar, and, by implication, to the BLR 
(Turnshek et al 1996). To some extent some of these abundances are 
model dependent (for example on a one- or two-component photoionisation 
model), so it remains to
be seen if they persist to the same extent in our stratified picture. 
If they do, then they imply either a large abundance gradient across 
the BLR, which may not be compatible with the emission line ratios, or that 
the BAL gas cannot be identified with the BEL gas and hence that BALs 
are a separate class. However, this would not solve the problem for us, 
since it would leave us with a class of edge-on objects that apparently 
has no observational counterpart. BALs therefore seem to provide a crucial 
test for the model we are proposing.

\section{Radio Loud Objects}

The infrared to soft X-ray continua, as well as the optical and UV properties
of lobe-dominated radio source and radio-quiet AGN are generally similar
(Sanders et al. 1989). This is a strong indication that they are produced 
in the same
way and that the quest for a unified picture of radio-loud and radio-quiet
objects is not a lost cause.        
The basic starting point for our unified model with regard to radio-loud sources
is therefore that these are the radio-quiet systems with a radio jet added in
each case. Thus we ignore, at least in the first instance, any 
interaction between the radio jet, or jet emission, and the other 
structures in the model. 

Of course, the properties of the radio emission itself may depend
on the radio luminosity, as indeed it does. We recognise two luminosity
classes (Fanaroff and Riley 1974): the edge-darkened FRI sources with 
radio power $<10^{25.3}$W Hz$^{-1}$ and the edge brightened FRII sources with
radio powers $>10^{25.3}$W Hz$^{-1}$. Since we are assuming the radio and BLR
properties are independent, this classification does not affect our
hypothesis that the radio quiet and radio loud systems 
should form parallel
series. Our point in this section is that the adoption of our series for the
radio-quiet systems overcomes some of the problems in achieving this
parallelism within the `SPM' unified scheme.

We adopt the angle dependence of the usual unified picture for 
radio-loud systems
(see Urry \& Padovani 1995 for a review) since the structure of the radio jet
gives a good indication of the angle to the line-of-sight.
However there should be present amongst radio-loud
systems the analogues of the `true' Seyfert 2s and the obscured Seyfert 1s. The
latter class has a representative in 3C 234 in which the broad lines  
were detected
in polarised light before those in NGC 1068 (Antonucci \& Cohen1983). 

Barthel (1989, 1991), 
in his unified scheme, stipulates that quasars and radio galaxies
have the same relation as Seyfert 1 and Seyfert 2 galaxies with the
central quasar in the narrow line radio galaxies 
obscured by an absorbing torus. One of the difficulties with Barthel's model
is this: Laing et al. (1993) studied a complete subsample
of 3CR radio sources with $z <$0.88 and found two classes of FRII
narrow-line spectra: low and high excitation.
Within the high-excitation
spectra they find comparable numbers of objects with and without detectable
broad lines at all redshifts. These fit into Barthel's unified
scheme. They detect no broad lines in the low-excitation
objects, but a few objects show strong radio cores and other indications 
of beaming. For the low-excitation group they therefore 
suggest a picture more reminiscent of the 
BL Lac/FR I unified scheme (Blandford and Rees 1978; Browne 1983; Antonucci 
and Ulvestad 1985; Urry et al. 1991).

Murphy et al. (1993) found that BL Lac objects are not more core-dominated 
than the quasars in a well-defined sample of powerful radio sources. 
They find no evidence that BL Lacs are those quasars seen at such small 
angles to the line of sight that the relativistically beamed core emission 
swamps that from other components, and come to the view that the majority 
of BL Lacs are related to the low-luminosity FRI radio galaxies. However, they
observed a substantial fraction of BL Lac objects with extended radio emission
with powers and structures more characteristic of FRIIs. These FRII-luminosity
BL Lacs are difficult to account for in the context of the standard unified
scheme (although if the radio axis were a function of time this could 
account for the extended emission). One of the suggestions of Murphy et al., 
namely that the BL Lacs know nothing of the FRI-FRII division and can be 
beamed versions of either type of radio source, fits into our picture. 

In our model the structure of the BLR is independent of radio power. This 
implies that the cut-off in cloud production in the BLR, that is, 
the classification of radio-loud objects with or without broad lines, 
must be independent of the transition between FRII and FRI objects.  
As a consequence we expect to observe either FRIIs without BLRs or FRIs 
with BLRs. Note that the model predicts an inner BLR radius 
$\propto T_{c}^{-1/3}$ (Cassidy \& Raine 1996) so if the steep 
X-ray spectrum implies a Compton temperature, $T_{c}$ as 
low as $10^{6}$K for the radiation falling on the disc then the BLR 
will be largely absent. The observation of relatively narrow 
broad wings to H$\alpha$ in a number of BL Lacs does not imply the 
existence of a BLR as these can arise in the ILR. (We expect that at 
least the higher luminosity FRIs/BL Lacs still posses their ILRs.) 
Hence FRII-luminosity BL Lacs could represent our FRIIs with absent 
BLRs and would be the radio-loud true Seyfert 2 analogues.

Another problem concerns the BAL QSOs: they appear to have no radio loud
analogues. Such analogues would be expected in a spherically symmetric picture 
where broad absorption lines from clouds in the line of sight,
are uncorrelated with the radio axis. 
But we do find Seyfert type absorption in radio loud systems. This
points to a picture in which, contrary to our initial hypothesis, 
the radio jet and disc structure are correlated to the extent that  
the presence of the radio jet rules out the
possibility of a flattened disc. Such a correlation would be entirely natural
(although not proven) in our picture. Recall that the disc flaring is 
suppressed at high luminosity (= high wind power) by the disc-wind 
interaction. In radio-loud systems it would be natural to suppose that  
the momentum carried away in the jet is
not available in the nuclear wind.  Hence the nuclear wind carries 
insufficient momentum to suppress the disc flaring.
So we expect to see some flaring of the accretion disc
even in the highest luminosity radio-loud objects. This is also indicated by
the existence of high luminosity NLRGs.
The disc flaring also settles the question of why the high luminosity blazars
and QSRs, which are viewed face-on, or close to face-on, do not show 
relatively narrow broad emission lines as would be expected by comparison with 
the face-on QSOs (section 5.1). 

Wills and Brotherton (1996) attributed the existing differences in the 
optical-UV spectra between radio-loud and radio-quiet objects to 
two apparently independent relations: (i) For radio-loud quasars, 
with increasing core dominance, emission lines are narrower, FeII emission 
is stronger , and CIV has stronger red wings. With decreasing 
core-dominance, H$\beta$ more often has stronger red wings, and the 
likelihood of associated absorption and reddening increases. 
(ii) For radio-quiet objects there is an inverse relation between the 
strengths of FeII and [OIII]$\lambda$ 5007 emission, with the strong 
FeII-weak [OIII] QSOs being associated with low ionisation BALs, 
reddening and polarisation.

We can explain these observations as follows: 

(i) the narrowing of the emission lines with increasing 
core-dominance is due to the disc-like distribution of the BLR 
clouds. CIV has stronger red wings because the CIV cloud profile 
is symmetric for face-on objects but the reflected component boosts 
the red wing. The increase in FeII emission can be explained if FeII 
arises in the accretion disc (e.g. Joly, 1991). That the likelihood 
of absorption and reddening increases with decreasing core-dominance 
is evident when considering Fig 1. At present we cannot explain why 
H$\beta$ more often has stronger red wings with decreasing core-dominance.

(ii) We would expect an increase in reddening and polarisation for 
edge-on objects (section 6.7). The weak [OIII]$\lambda$5007 emission in 
low ionisation BAL QSOs and the strong FeII emission from edge-on radio-quiet 
objects  (contrary to the trend in radio-loud quasars) is more difficult 
to explain. However, weak [OIII] is not sufficient to predict the 
presence of BALs; there are examples of strong [OIII] emitters 
amongst BAL QSOs (Turnshek 1995). We can speculate that the difference 
in FeII is associated with the different structure of the accretion discs.
 
In figure 4(b) we show the radio-loud analogues of QSOs and Seyferts at high and
low luminosity and their classification when viewed from various angles.  
For completeness we summarise briefly here what this picture shows, although
none of the following is new. 
At the high luminosity end we have OVVs and 
HPQs, radio-loud 
quasars and FRIIs. OVVs and HPQs are observed when we look directly at the jet.
At larger cone-opening angles we observe flat-spectrum (core-dominated) 
and steep-spectrum (lobe-dominated) QSRs. At even larger angles we
have broad-line radio-galaxies. Objects where the BLR or BLR and ILR are
obscured by the disc are narrow-line radio galaxies (NLRGs). 

At the lower luminosity end we have BL Lacs and FR I radio galaxies.
The idea is that FR I radio galaxies have relativistic jets (this is not
yet proven)
in which nonthermal emission, when viewed from small angles, dominates
the emission of the galaxy. They are the parent population of
BL Lacs (Blandford \& Rees 1978, Blandford \& K\"{o}nigl 1979, Orr \&
Browne et al. 1982, Browne 1983, Wardle et al. 1984, 
Antonucci \& Ulvestad 1985, Sarazin \& Wise 1993). At even lower luminosities
we have radio-loud LINERS.
Most of the AGN-type LINERs found in bright, nearby galaxies are associated with 
weak compact radio sources; however, some LINERs have been found in 
the nuclei of powerful radio galaxies, eg. M87 (Goodrich \& Keel 1986).             

An interesting suggestion was made by Falcke et al. (1995, 1996). They 
stipulate that the flat spectrum radio intermediate quasars are actually 
boosted radio-quiet objects. This implies the assumption that radio-quiet 
objects also have relativistic jet, but a factor of  $\sim$ 100-1000 less 
luminous than the radio-loud objects. If this is correct then flat 
spectrum quasars should appear in figure 4a as face-on QSOs and not 
in fig. 4b. Our model then predicts that these objects will have 
relatively narrow broad emission lines.                      
                                      
\section{Conclusions}

We begin by summarising the way in which the picture we have presented in this 
paper differs from the molecular torus unified model. The torus at large radii 
obscures the
direct view of the BLR and material from the torus provides the 
scattering medium by which the BLR
is indirectly visible. The torus is also responsible for the beaming of the UV.
There is a fixed cone opening angle which is not in agreement with observation. 
Modifications to this  must be made ad hoc.

In our picture the BLR clouds themselves sustain the scattering medium.
Except at high luminosity (assuming this implies a high mass outflow) 
the disc flares in the BLR, partially collimating the nuclear outflow away from
the disc plane. At larger radii a ring of dusty material (possibly
associated with the disc) then provides an
obscuring screen. At high luminosity the disc flaring is suppressed and the
presence of a strong wind close to  the disc plane removes the obscuring
ring (leaving a geometrically thin, opaque, dusty disc).
The angle subtended by the 
scattering cone will in general be different from the UV cone.
The obscuring cone depends on luminosity in a predictable way, through
the disc flaring, in
general qualitative agreement with the data. The model for the different
luminosity states of NGC 4151 at a fixed angle can be extended 
to describe other active nuclei of
various luminosity classes at different angles. 

A unified model based on NGC 4151 as archetype leads to the
following conclusions. (But note we are not claiming it is the 
only model that can do so.)

\begin{enumerate}
\item Principally as a result of  
the non-spherical BLR geometry 
\begin{enumerate}
\item there is no conflict between the
Hubble data and the alignment of the UV cone and radio structure in NGC 4151; 
\item a range of X-ray columns
are possible with, depending on the line of sight, 
partial covering in some cases
and warm absorbers in others  (and, in the case of NGC 4151, both). 
\item there are no BAL Seyferts or radio loud BAL quasars.
\end{enumerate}                                                            
\item As a consequence of a luminosity dependent BLR, (possible because the ram
pressure accelerated clouds are short-lived):
\begin{enumerate}
\item there are nuclei that appear (for an inflow timescale, $> 10^{5}$ years) as 
true Seyfert 2s with low X-ray absorption;    
\item classification changes
in Seyferts are possible on short time scales (weeks to months). 
\end{enumerate}
\item  The luminosity dependence of the disc flaring and obscuration 
means that the picture is compatible with                   
\begin{enumerate}
\item the lack of evidence
for obscuring dust in some Seyfert galaxies 
(or positive evidence that it is not there); 
\item the possibility that there are no high luminosity Seyfert 2s 
(assuming that the IRAS galaxies are young QSOs that have recently switched on);
\item variations in opening angle of the obscuring torus, which are implied by 
observations, and also required if Seyfert galaxies are not to overproduce
the X-ray background (Mulchaey et al. 1992).
\end{enumerate}                                                          
\item The contribution of the ILR to the `narrow' line emission, implies that        
\begin{enumerate}
\item there is no missing [OIII] $\lambda 5007$ problem in some NLRGs and
BLRGs; 
\item the NLR profiles in galaxies classified observationally as lower 
luminosity Sy2s, should show no evidence for any ILR component. This is at
least consistent with the similarity of profiles for different lines within
these systems.
\item the problem of  assigning a consistent temperature from [OIII] $\lambda
4363$ in photoionisation models is resolved.
\item the [OIII] $\lambda 5007$ line can have a blue wing without the need to
invoke NLRs with high densities. 
\end{enumerate}
\item Finally, the combination of BLR geometry and short-lived BLR clouds means
\begin{enumerate}
\item the NLR cone has a larger angular extent 
than the scattering cone (which is a conical shell).
\end{enumerate}
\end{enumerate}

NGC 1068 can be fitted into this picture because the exceptionally large column
of dust it contains is special to its line of sight. (It may be associated
with the equatorial plane of a star formation ring in the galaxy.)
The model predicts a correlation between X-ray absorption and 
both high-ionisation forbidden
line profiles and broad-line profiles, and a correlation between UV and X-ray
absorption at certain angles. There should also be  a 
relation between the luminosity function and profile
width distribution for radio-quiet objects. Broad-line radio galaxies 
appear over a different range of
angles from radio quasars and so should have broader lines on average. 
These may provide possible tests which we hope to discuss
elsewhere.\\                 
\\
Acknowledgement: We are grateful to Dave Smith for directing us to the 
X-ray data of figure 2 and guiding us in the use of XSPEC to analyse it, 
and to two anonymous referees for their constructive scholarship. \\
\\
                                                                               
\noindent
\Large
{\bf References}\\
\normalsize
\\
\noindent
 Alloin, D., Pelat, D., Phillips, M., Whittle, M., 1985, Astrophys. J., 288, 
 205 \\
 Andrillat, Y. \& Souffrin, S., 1968, Astrophys. Lett., 1, 111 \\
 Antonucci, R.R.J., Cohen, R.D., 1983, Astrophys. J., 271, 564 \\
 Antonucci, R.R.J., 1984, Astrophys. J., 278, 449    \\
 Antonucci, R.R.J., 1993, Ann. Rev. of Astron. Astrophys., 31, 473 \\
 Antonucci, R.R.J., Ulvestad, J.S., 1985, Astrophys. J., 294, 158 \\
 Antonucci, R.R.J., Hurt, T. \& Miller, J., 1994, Astrophys.J., 430,210 \\ 
 Arav, N., Korista, K.T., Barlow T.A. \& Begelman M.C., 1995, Nature, 376, 576\\ 
 Baker, J.C. \& Hunstead, R.W.,   1995   Astrophys. J. 452, L95\\
 Barr, P., Willis, A.J. \& Wilson, R., 1983, MNRAS, 203, 201\\
 Barthel, P.D., 1989, Astrophys. J., 336, 606 \\
 Barthel, P.D., 1991, Physics of Active Galactic Nuclei, eds. W.J. Duschl and
 J.S. Wagner, Springer Verlag, p. 637 \\
 Begelman, M.C., McKee, C.F., Shields, G.A., 1983, Astrophys. J. 271, 70\\
 Blanco, P.R., Ward, M.J., Wright, G.S., 1990, MNRAS, 242, 4P\\
 Blandford, R.D. \& McKee, C.F., 1982, Astrophys.J., 255, 419\\
 Blandford, R.D., Rees, M.J., 1978, in Pittsburgh Conference on BL Lac
 Objects, p. 328, ed. A. Wolfe \\
 Blandford, R.D. \& K\"{o}nigl, A., 1979, Astrophys.J., 232, 34 \\
 Boller, T., Brandt, W.N. \& Fink H.,  1996, Astron \& Astrophys., 305, 53\\
 Boyle, B.J.  \& di Malteo, T., 1995, MNRAS, preprint\\
 Bregman, J.N., 1991, Physics of Active Galactic Nuclei, eds. W.J. Duschl and
 S.J. Wagner, Springer Verlag, p. 699 \\
 Bromage et al., 1985, MNRAS, 215,1\\
 Brotherton M.S., Wills, B.J., Francis, P.J. \& Stendal, C.C., 1994, 
Astrophys.J., 430, 495 \\ 
 Brotherton, M.S., 1996, Astrophys. J. Supp., 102, 1\\
 Browne, I.W.A., et al., 1982, MNRAS, 198, 673 \\
 Browne, I.W.A., 1983, MNRAS, 204, 23P \\
 Capetti, A., Axon, D.J., Maschetto, F., Sparks, W.B. \& Boksenberg, A.,1995, 
Astrophs.J., 446, 155  \\
 Cassidy, I., Raine, D.J., 1993, MNRAS, 260, 385    \\
 Cassidy, I., Raine, D.J., 1996, Astron. Astrophys. 310, 48\\
 Clement, R., Sembay, S., Hanson, C.G., Coe, J.M., 1988, MNRAS, 230, 117 \\
 Cohen, M.H., Ogle, P.M., Tran, H.D. Vermeulen, R.C. 1995, Astrophys.J. 448,
L77\\
 Crenshaw, D.M., Peterson, B.M. \& Wagner, R.M., 1988, Astron.J., 96, 1208
 de Robertis, M., 1987, Astrophys. J., 316, 597  \\
 Done, C. \& Krolik, J.H.,  1996, Astrophys. J., 463, 144  \\
 Edelson, R.A. \& Malkan, M.A., 1986, Astrophys.J., 308, 59\\
 Efstathiou, A. \& Rowan-Robinson, M., 1995, MNRAS, 273,649\\
 Evans, I.N., Tsvetano, Z., Kriss, G.A., Ford, H.C., Caganoff, S. \& Koratkar,
A.C., 1993, Astrophys.J., 417, 82\\
 Fahey, R.P., Michalitsianos, A.G., Kazanas, D., 1991, 
 Astrophys. J. 371, 136 \\
 Falcke, H., Gopal-Krishna, Biermann, P.L.,  1995, Astron. Astrophys., 298, 395\\
 Falcke, H., Sherwood, W., Patnaik, A.R.,  1996,  preprint \\
 Fanaroff, B.L., Riley, J.A., 1974, MNRAS 167, 31P \\
 Ferland, G.J., Netzer, H., 1983, Astrophys. J., 264, 105 \\
 Ferland, G.J., Peterson, B.H., Horne, K., Welsh, W.F. \& Nahar, S.N., 1992,
Astrophys.J., 387, 95\\
 Filippenko, A.V., 1985, Astrophys. J., 289, 475 \\
 Filippenko, A.V., Halpern, J.P., 1984, Astrophys. J., 285, 458  \\
 Filippenko, A.V., Sargent, W.C.W., 1986, Observational Evidence of Activity
 in Galaxies, eds. E.Ye. Khachikian, K.J. Fricke \& J. Melmick, p. 451 \\
 Francis, P.J., Hewett, P.C., Foltz, C,B. \& Chaffee, F.H. 1992, 
Astrophys. J., 398, 476\\
 Giannuzzo, E., Rieke, G.H., Rieke, M.J., 1995, Astrophys.J., 446, L5 \\
 Giuricin, G., Mardirossian, F., Mezzetti, M., 1995, Astrophys. J. 446, 550\\
 Goodrich, R.W. \& Keel, W.C., 1986, Astrophys. J., 305, 148\\
 Goodrich, R.W. Miller, J.S. 1995, Astrophys.J., 448, L73\\
 Grandi, S.A., 1978, Astrophys. J., 221, 501\\
 Greenhouse, M.A., Feldman, U., Smith, H.A., Klapisch, M., Bhatia, 
A.K. \& Barshalom, A., 1993, Astrophys. J. Supp.88, 23\\
 Halpern, J.P., Steiner, J.E., 1983, Astrophys. J., 269, L37 \\
 Heckman, T.M., 1980, Astron. Astrophys., 87, 142 \\
 Heckman, T.M., Miley, G.K., Green, R.F., 1984, Astrophys. J., 281, 525 \\
 Heckman, T.M.,1986, Observational Evidence for Activity in AGN, eds. E.Ye.
 Khachikian et al., p. 421 \\
 Heckman, T.M., Chambers, K., Schilizzi, R. \& Postman, M., 1992, 
Astrophys.J., 391,39\\
 Heckman, T.M., O'Dea, C.P., Baum, S.A. \& Laurikainen, E.,  1994, 
Astrophys. J., 428, 65 \\
 Heckman, T.M. et al., 1995, Astrophys. J., 446, 101 \\
 Hes, R., Barthel, P.D., Fosbury, R.A.E., 1993, Nature, 362, 326 \\
 Hutchings, J.P., 1987, Astrophys. J., 320, 122 \\
 Inda, M., Makishima, K., Kohmura, Y., Tahsiro, M., Ohashi, T., Barr, P.,
 Hayshida, H., Palumbo, G.G.C., Trinchieri, G., Elvis, M., Fabbiano, G.,
 1994, Astrophys. J., 420, 143\\
 Jijima, T.,  1992, IAU Circ., 5521 \\
 Jijima, T., Rafanelli, P., Bianchini, A., 1992, Astron. Astrophys., 265, L25 \\
 Joly, M., 1991, Astron. Astrophys. 242, 49\\
 Kay, L.E.,  1994, Astrophys. J., 430, 196\\
 Kinney, A.L., Antonucci, R.R.J., Ward, M.J., Wilson, A.S., Whittle, M.,
 1991, Astrophys. J., 377, 100 \\
 Kollatschny, Fricke, 1985, Astron. Astrophys., 146, L11 \\
 Kwan, J., 1984, Astrophys. J. 283, 70 \\
 Laing, R.A., Jenkins, C.R., Walls, J.V., Unger, S.W., 1993, Poster Paper
 presented at Mt. Stromlo Symposium in June 1993 \\
 Lawrence, A., 1991, MNRAS 252, 586 \\
 Loska, Z., Czerny, B., Szczerba, R., 1993, MNRAS 262, L31 \\
 Maiolino, R., Ruiz, M., Rieke, G.H. \& Keller, L.D., 1995, 
Astrophys.J., 446, 561\\
 Maisack, M., Yaqoob, T., 1991, Astron. Astrophys., 249, 25 \\
 Maoz, D., et al. 1991, Astrophys.J., 367, 493 \\
 Marconi, A., Moorwood, A.F.M., Salvat, M., Oliva, E., 1994, Astron. 
Astrophys, 291, 18\\
 Marshall, F.E. et al., 1993, Astrophys.J., 405, 168 \\
 Mathews, W.G. \& Ferland, G.J., 1987, Astrophys. J., 323, 456\\
 Mathews, W.G., 1993, Astrophys.J., 412, L17\\
 McHardy, I. \& Czerny, B., 1987, Nature, 325, 696\\
 Mobasher, B., Raine, D.J., 1989, MNRAS 237, 979 \\
 Mulchaey, J.S., Mushotzky, R.F., Weaver, K.A., 1992, Astrophys. J., 
390, L69 \\
 Murphy, D.W., Browne, I.W.A., Perley, R.A., 1993, MNRAS, 264, 298\\
 Nandra, K., Pounds, K.A., Stewart, G.C., Fabian, A.C. \& Rees M.J., 1989, 
MNRAS, 236, 39\\
 Nandra, K. \& Pounds, K.A., 1994, MNRAS, 268, 405 \\
 Netzer, H., 1990, Active Galactic Nuclei, p. 57, Swiss Society for Astrophysics
 and Astronomy, eds. R.D. Blandford, H. Netzer, L. Woltjer, Springer Verlag \\
 Nussbaumer, H., \& Osterbrock, D.E., 1970, Astrophys. J., 161, 811\\
 Oke, J.B. \& Sargeant, W.L.W., 1968, Astrophys. J., 151, 807\\
 Oke, J.B., Redhead, A.C.S. \& Sargeant, W.L.W., 1980, PASP, 92, 758\\
 Oliva, E. \& Moorwood, A.,F.,M., 1990, Astrophys. J. 348, L5\\
 Oliva, E., Salvati, M., Moorwood, A.F.M. \& Marconi, A., 1994, Astron. 
Astrophys., 288, 457\\
 Orr, M.J.L., Browne I., 1982, MNRAS 200, 2067 \\
 Pan, H.C., Stewart, G.C. \& Pounds, K.A., 1990, MNRAS, 242, 177\\
 Pedlar, A., Howley, P., Axon, D. J. \& Unger, S.W., 1992, MNRAS, 259, 369\\  
 Penfold, J., 1992, IAU Circ., 5533 \\
 Penfold, J., 1992a, IAU Circ., 5547 \\
 Penston, M.V., et al. 1990, Astron. Astrophys. 236, 53\\
 Penston, M.V., Perez, E., 1984, MNRAS, 211, 33P \\
 Pogge, R.W., 1989, Astrophys. J., 345, 730 \\
 Pounds, K.A., Nandra, K., Stewart, G.C. \& Leighley, K., 1989, MNRAS, 240,
769\\
 Pounds, K.A., Nandra, K., Stewart, G.C., George, I.M., Fabian, A.C., 1990
 Nature, 344, 132 \\
 Puchnarewicz, E.M., et al. 1992, MNRAS, 256, 589\\
 Reichert, G.A., Mushotzky, R.F., Petre, R., Holt, S.S., 1985, Astrophys. J.,
 296, 69 \\
 Reynolds, C.S. \&  Fabian A.C., 1995, MNRAS, 273, 1167\\
 Robson, E.I. et al. 1986, Nature, 323, 134 \\
 Rodriguez Espinosa, J.M., Rudy, R.J., Jones, B., 1986, Astrophys. J., 
 309, 76 \\
 Rodriguez Espinosa, J.M., Rudy, R.J., Jones, B., 1987, Astrophys. J., 
 312, 555 \\
 Sanders,D.B., Phinney, E.S., Neugebauer, G., Soifer, B.T., Mathews, K,
 1989, Astrophys. J., 347, 29 \\
 Sarazin, C.L., Wise, M.W., 1993, Astrophys. J., 411, 55 \\
 Shuder J.M., 1980, Astrophys. J., 240, 32 \\
 Smith, D.A. \& Done, C., 1995, MNRAS (in press)\\
 Smith. M,J. \& Raine, D.J., 1985, MNRAS, 212, 425\\
 Smith, M.J \& Raine, D.J., 1988, MNRAS, 234, 297 \\
 Spinoglio, L..\& Malkan, M.A.,  1992, Astrophys. J., 399, 504\\
 Stephens, S.A., 1989, Astron. J., 97, 10\\
 Stirpe, G.M., van Gr\"{o}ningen, E., de Bruyn A.G., 1989, Astron. 
Astrophys. 211, 310\\
 Storchi-Bergmann, T., Baldwin, J.A., Wilson, A.S., 1993, Astrophys. J.,
 410, 111 \\
 Tadhunter, C. \& Tsvetanov, Z., 1989, Nature, 341, 422\\
 Telesco et al  1984, Astrophys.J., 282, 428\\
 Thompson, R.I., 1995, Astrophys.J., 445, 700\\
 Tran, H.D., 1995, Astrophys. J. 440, 565\\
 Turnshek, D.A., 1987, Proceedings of the QSO Absorption Line Meeting, 
 Baltimore, eds. J.C. Blades, D.A. Turnshek, C.H. Norman, p.1 \\
 Turnshek, D.A., 1995, in {\em QSO Absorption Lines} ed. G.Meylan, 
(Springer) 223\\
 Turnsheck, D.A., Kopke, M., Monier, E., Noll, D., Epsey, B.R. 
\& Weymann, R.J., 1996, Astrophys. J., 463, 110\\
 Ulrich, M.-H., et al., 1984, Astrophys. J., 382, 483   \\
 Ulrich, M.-H., 1988, MNRAS, 230, 121\\ 
 Ulrich, M.-H. \& Horne, K., 1996, preprint, MNRAS\\
 Ulvestad, J.S., 1986, Astrophys. J., 310, 136 \\
 Unger, S.W., Taylor, K., Pedlar, A., 1987, Active Galactic Nuclei, eds. 
 D. Osterbrock, Miller, IAU Symp. 134, Kluwer Verlag, 331  \\
 Urry, C.M., Padovani, P., Stickel, M., 1991, Astrophys. J., 382, 501 \\
 Urry, C.M. \& Padovani, P., 1995, PASP, 107, 803\\
 van Gr\"{o}ningen, E.  \& be Bruyn, A. G.,1989, Astron. 
Astrophys., 211, 293 \\
 Viegas-Aldrovandi S., M. \& Contini, M., 1989  Astron. Astrophys. 215, 253\\
 Voit, G.M., Weymann, R.J., Korista, K.T., 1993, Astrophys. J., 413, 95 \\
 Wamsteker, W., et al., 1985, Astrophys. J., 295, L33 and L45 \\
 Wamsteker, W.  et al., 1990, Astrophys.J., 354, 446 \\
 Ward, M.J., et al., 1978, Astrophys. J., 223, 788 \\
 Warwick, R.S., Done, C. \& Smith, D.A., 1995, MNRAS, 275, 1093\\
Wardle, J.F.C., Moore, R.L., Angel, J.R.P., 1984, Astrophys. J., 279, 93 \\
 Weaver et al., 1994a, Astophys. J., 423, 621 \\
 Weaver et al., 1994b, Astrophys. J. 436, L27\\
 Webster, R.L., Francis, P.J., Peterson, B.A., Drinkwater, H.J., 
Masci, F.J., 1995, Nature, 375, 469\\
 Weymann, R.J., Turnshek, D.A.\& Christianson, W.A., 1985, 
in {\em Astrophysics of Active Galaxies and QSOs} ed. J.S.Miller, p333\\
 Weymann, R.J., Morris, S.L., Foltz, C.B., Hewett, P.C., 1991,
 Astrophys. J., 373, 23 \\
 Whittle, M., 1985, MNRAS, 213, 1 and 33 \\
 Whittle, M., 1988, MNRAS, 216, 817 \\
 Wills, B.J., Brotherton, M.S. \& Fang, D., 1993, Astrophys. J., 415, 563\\
 Wills, B.J. \& Brotherton, M.S., 1996, preprint\\
 Wilson, A.S., Heckman, T.M., 1985, Astrophysics of Active Galaxies and QSOs,
 ed. J.S. Miller, p. 39 \\
 Wilson, A., Ward, M. \& Haniff, C., 1988, Astrophys. J., 334, 121\\
 Wilson, A., 1991, Physics of Active Galactic Nuclei, eds. W.J. Duschl,
 S.J. Wagner, Springer Verlag, p. 307 \\
 Woltjer, L., 1990, Active Galactic Nuclei, p. 1, Swiss Society for 
Astrophysics  and Astronomy, eds. R.D. Blandford, H. Netzer, L. Woltjer, 
Springer Verlag \\
 Yaqoob, T. \& Warwick, 1991, MNRAS, 248, 773\\
 Yaqoob, T. Warwick, R. S., Makino, F., Otani, C., Sokoloski, J.L., 
Bond, I.A., \&      Yamauchi, M., 1993, MNRAS, 262, 435 \\
 Yee, H.K.C. \& Oke, J.B., 1981, Astrophys.J., 248, 472\\
 Zheng, W., 1996, Astron. J. preprint\\

\newpage
\Large
\noindent {\bf Figure captions}\\
\normalsize
\\
{\bf Figure 1} shows the structure of NGC 4151 including the warm absorber and
reflector inferred from X-ray studies. 
We have a flared accretion disc with an optically thick, geometrically
thin layer surrounding a black hole with either strong or weak radio jets.
The flaring and thickness of the disc are, of course, luminosity dependent.
The X-rays are created close to the black hole and are reflected by part of 
the inner accretion disc (Nandra et al. 1989). At distances $\geq 10^{16}$ cm 
broad-line clouds are formed. These clouds break up within two 
sound-crossing times and the cloud remnants form the warm absorber (WA). 
At distances $\geq 10^{18}$ cm the intermediate emission line clouds form. 
At distances $\geq 10^{19}$ cm
we find the narrow line clouds, which are probably distributed 
isotropically.\\
\\
{\bf Fig 2.}   X-ray counts and folded model and $\chi^{2}$ versus energy 
for Ginga observations of NGC 4151 for (a) May 1990
and (b) Nov 1990. \\
\\
{\bf Fig. 3.} Transfer functions for the
disc-wind model at two disc inclinations (a) (dashed line) 24$^{\circ}$ and 
(b) (solid line) 75$^{\circ}$.\\
\\               
{\bf Fig. 4} (a) High and low luminosity radio-quiet AGN as 
seen from various lines of sight. (b) High and low luminosity radio-loud 
AGN as seen 
from various lines of sight.


\newpage
\begin{minipage}[t]{7in}
\begin{tabular}{lllllll}
\multicolumn{3}{c}{TABLE OF MODEL PARAMETERS} \\
\multicolumn{3}{l}{ } \\
Symbol & Quantity & Value in NGC 4151 \\
\small
$L_{r}$ & bolometric luminosity & $3\times 10^{43}$ erg s$^{-1}$\\
$L_{E}$ & Eddington limit luminosity &\\
$M_{bh}$& black hole mass& $3\times 10^{7} {\rm M}_{\odot}$\\
$L_{m}$ & momentum flux in wind& ....\\
$r_{in}$& inner radius of BLR& 1.4$\times 10^{16}$ cm\\
$r_{out}$& outer radius of BLR & $10^{17}$ cm\\
$v_{n}$& cloud injection speed normal to the disc & $2\times 10^{8}$ cm s$^{-1}$\\
$v_{\phi}$ & azimuthal cloud velocity &  \\
$v_{\phi0}$ & initial value of $v_{\phi}$& Kepler speed at input radius\\
$t_{s}$ & sound crossing time in cloud at formation &$5\times 10^{3}r_{16}^{2}$ s \\
$t_{c}$ & cloud survival time & $2t_{s}$\\
$A_{d}$ & disc albedo & 0.3\\
$n$ & cloud density & \\
$N_{c}$ & column density in clouds & $10^{23}/(1+(t/t_{s})^{2})$ cm$^{-2}$\\

\end{tabular}
\end{minipage}

\end{document}